\documentclass{article}

\usepackage[margin=1in]{geometry}

\usepackage{amssymb}

\usepackage{appendix}
\usepackage{amsmath}
\usepackage{graphicx}
\usepackage{lineno}
\usepackage{array}
\usepackage{longtable}
\usepackage[numbers,sort&compress]{natbib}
\usepackage{caption}
\usepackage{subcaption}
\usepackage{amsthm}
\usepackage{amsmath,amsfonts,amssymb}
\usepackage{graphicx}
\usepackage{setspace}
\usepackage{tocloft}
\usepackage[colorlinks=True, linkcolor=blue, citecolor=blue, urlcolor=blue]{hyperref}
\usepackage{bm}
\usepackage{fancyhdr}
\usepackage{mathtools}
\usepackage{algorithm}
\usepackage{algpseudocode}
\usepackage{comment}
\usepackage{authblk}
\newcommand{\eleint}[1]{\int_{\Omega_e} {#1}\, d\Omega_e}

\newcommand{\upoly}{\bm{u_p}}
\newcommand{\bpsi}{\bm{\psi}}

\newtheorem{theorem}{Theorem}[section]
\newtheorem{lemma}[theorem]{Lemma}

\newcommand{\meansize}{0.7}
\newcommand{\rmssize}{0.45}

\newcommand*\patchAmsMathEnvironmentForLineno[1]{%
\expandafter\let\csname old#1\expandafter\endcsname\csname #1\endcsname
\expandafter\let\csname oldend#1\expandafter\endcsname\csname end#1\endcsname
\renewenvironment{#1}%
{\linenomath\csname old#1\endcsname}%
{\csname oldend#1\endcsname\endlinenomath}}%
\newcommand*\patchBothAmsMathEnvironmentsForLineno[1]{%
\patchAmsMathEnvironmentForLineno{#1}%
\patchAmsMathEnvironmentForLineno{#1*}}%
\AtBeginDocument{%
\patchBothAmsMathEnvironmentsForLineno{equation}%
\patchBothAmsMathEnvironmentsForLineno{align}%
\patchBothAmsMathEnvironmentsForLineno{flalign}%
\patchBothAmsMathEnvironmentsForLineno{alignat}%
\patchBothAmsMathEnvironmentsForLineno{gather}%
\patchBothAmsMathEnvironmentsForLineno{multline}%
}

\begin{document}
\title{A Spectral Element Enrichment Wall-Model}
\date{November 15, 2024}
\author[1,2,*]{Steven R. Brill}
\author[3]{Pinaki Pal}
\author[3]{Muhsin Ameen}
\author[3]{Chao Xu}
\author[4]{Matthias Ihme}

\renewcommand\Affilfont{\fontsize{9}{10.8}\itshape}
\affil[1]{Institute for Computational and Mathematical Engineering, Stanford University, Stanford, CA}

\affil[2]{Design Physics Division, Lawrence Livermore National Laboratory, Livermore, CA}

\affil[3]{Transportation and Power Systems Division, Argonne National Laboratory, Lemont, IL}

\affil[4]{Department of Mechanical Engineering, Stanford University, Stanford, CA}
\maketitle

\begin{abstract}
In the present work, a first-of-its-kind enrichment wall-model is developed within the spectral element method (SEM) framework for large-eddy simulations of wall-bounded turbulent flows. The method augments the polynomial solution in the wall-adjacent elements with an analytical law-of-the-wall enrichment function representing the mean velocity near the wall. In the solution representation, this enrichment function captures the large gradients in the boundary layer, which allows the polynomial modes to represent the turbulent fluctuations. The enriched solution is able to resolve the shear stress at the wall without any modification to the no-slip wall boundary conditions, which allows for greater accuracy in the near-wall region compared to traditional methods. The enrichment wall-modeling approach is implemented in a high-order SEM computational fluid dynamics (CFD) solver, Nek5000, and its performance is assessed in turbulent channel flow large-eddy simulations (LES) for a range of Reynolds numbers. It is demonstrated that the enrichment wall-model improves solution accuracy on under-resolved near-wall grids as compared to the traditional shear stress wall-models.
\end{abstract}









\section{Introduction}
High Reynolds number turbulent flows are a norm, not the exception in engineering systems. Correctly simulating wall-bounded turbulent flows is crucial for applications, such as turbomachinery and propulsion/power engines \cite{tucker2013trends}. In such applications, accurately resolving the velocity and thermal boundary layers at the wall is critical to compute the quantities of interest. However, prohibitive costs render computational fluid dynamics (CFD) simulations of these full-scale systems infeasible because they need to resolve the smallest turbulent eddies. 

For most industry-relevant flow problems, this computational cost issue is primarily addressed by using Reynolds-Averaged Navier-Stokes (RANS) simulations wherein all turbulent length and time scales are modeled and the mean flow solution is computed. For some problems, these methods are sufficient, but for unsteady simulations, simulations with flow separation, and flows with laminar-to-turbulent transition, RANS methods are unreliable. For such problems, the turbulence scales must be resolved throughout the domain in direct numerical simulations (DNS) or in the near-wall regions and modeled elsewhere in wall-resolved large-eddy simulations (WRLES). However, both of these approaches are often computationally infeasible because the number of grid points in the near-wall region scales with $Re^{2.05}$ for DNS and $Re^{1.86}$ for WRLES \cite{yang2021grid}. Only recently have LES methods begun to see use for industrial applications \cite{goc2021large,jain2020massively,bergmann2024numerical,angel2024immersed}. Hence, there is a need to develop reliable methods to reduce the cost of such simulations by modeling the smaller turbulent scales in the near-wall regions, but resolving the large-scale transients via wall-modeled large eddy simulations (WMLES).

There have been a number of different WMLES algorithms developed in the literature. Most have been developed for low-order finite-difference (FD) and finite-volume (FV) methods, \cite{larsson2016large} but only recently there have been efforts to develop such methods in the context of higher-order methods such as the Spectral Element Method (SEM) \cite{pal2021development}, Spectral Difference Method (SDM) \cite{lodato2014structural}, and the discontinuous Galerkin (DG) method \cite{frere2017application,carton2017assessment,lv2021discontinuous}. Most of these methods follow the tradition of low-order FV methods and use the solution outside of the near-wall region to fit an empirical wall function in the boundary layer in order to compute the wall shear stress. Then, this shear stress is applied to the flow as the wall boundary condition. While these methods work well away from the wall, the modified boundary condition leads to inaccuracies in the near-wall region.

This study seeks to make use of the high-order polynomial solution of SEM to develop a different kind of wall-model better suited for high-order methods than traditional models. The wall-model builds upon prior work on the development of so-called enriched basis turbulence wall models in the context of DG \cite{lv2021discontinuous,brill2020enriched} and the continuous finite element (FEM) \cite{krank2016new,krank2018wall,krank2019multiscale}. These methods leverage the polynomial solution and augment it by considering the empirical wall function as part of the solution representation. Doing so prevents spurious oscillations in the near-wall region due to insufficient polynomial resolution of the boundary layer and improves the solution representation by allowing the degrees of freedom to represent the flow fluctuations instead of the mean flow behavior. This work serves to develop the framework for enrichment wall-models in SEM so that SEM because of its low dissipation and dispersion errors for its cost \cite{wang2013high} and its computational efficiency and scalability \cite{offermans2016strong}.

In the present work, a first-of-its-kind enriched basis WMLES approach for high-order SEM is developed and its performance is compared against the traditional equilibrium wall-stress modeling framework. The method leverages the high-order SEM framework in the wall-model to improve the accuracy of the WMLES and prevent unphysical oscillations, and is easily implementable in existing SEM solvers. Specifically, this method is implemented in the high-order SEM CFD solver Nek5000 \cite{nek5000-web-page}. The method is validated in the context of WMLES calculations for test problems pertaining to turbulent channel flows for a range of Reynolds number.

\section{Governing Equations}
We consider incompressible, Newtonian flows as governed by the Navier-Stokes equations:
\begin{align}
    \frac{\partial \bm{u}}{\partial t} &= -\nabla p + \nabla \cdot \left[\nu\left( \nabla\bm{u}+\nabla\bm{u}^T\right)\right] - \bm{u}\cdot \nabla \bm{u} \label{eqn:NS} \\ 
\nabla \cdot \bm{u} &= 0 \label{eqn:cont}
\end{align}
where $\bm{u}$ is the velocity vector, $t$ corresponds to time, $p$ is the pressure, and $\nu$ is kinematic viscosity of the fluid. The equations are solved over a computational domain $\Omega$ with boundary $\partial \Omega$. In this work, we consider implicit LES (iLES), where the numerical dissipation of the scheme is used to represent the small scales, instead of an explicit subgrid-scale turbulence model.

\section{Spectral Element Method}
The governing equations are discretized using the spectral element method (SEM) in the Nek5000 solver\cite{nek5000-web-page}.  In the SEM, the governing equations are put into a weighted residual formulation by multiplying Eqs.~\eqref{eqn:NS}-\eqref{eqn:cont} by test functions $\bm{v}$ and $q$, respectively, and integrating over the elements, so the goal is to find $(\bm{u},p)\in X_b^P(\Omega)\times Y^P(\Omega)$ such that 
\begin{align}
	\frac{d}{dt}(\bm{v},\bm{u}) &= -\left(\bm{v}, \frac{1}{\rho}\nabla p\right) + (\bm{v},\nabla\cdot\nu(\nabla\bm{u}+\nabla\bm{u}^T)) - (\bm{v},\bm{u}\cdot\nabla\bm{u}) , \label{eqn:wr_ns1}\\
	(\nabla \cdot \bm{u}, q)&=0 , \label{eqn:wr_cont1}
\end{align}
for all test functions $(\bm{v},q)\in X_0^P(\Omega)\times Y^P(\Omega)$, where the $\mathcal{L}^2$ inner product is defined as $(\bm{f},\bm{g})\coloneqq \int_{\Omega} \bm{f}\cdot\bm{g}\, dV$. Here, $X^P(\Omega)\subset H^1(\Omega)$ is the set of continuous $P$th-order spectral element basis functions. $X^P_b$ is the subset of $X^P$ such that Dirichlet boundary conditions are satisfied on the boundary, $\partial \Omega$. The subset satisfying homogeneous Dirichlet boundary conditions is $X^P_0$. Similarly, $Y^P(\Omega)\subset H^1(\Omega)$ is the set of continuous $P$th-order spectral element basis functions \cite{deville2002high}. The specifics of these basis functions will be defined later in this section, because the development of the weak form does not depend on the specific basis. From here, Eqs.~\eqref{eqn:wr_ns1}-\eqref{eqn:wr_cont1} are integrated by parts to reduce the derivative on $\bm{u}$ and create a symmetric form. This gives the final weighted residual form:

\begin{align}
    \frac{d}{dt}(\bm{v},\bm{u}) &= \left(\nabla \cdot \bm{v}, \frac{1}{\rho}p\right) - (\nabla\bm{v},\nu(\nabla\bm{u}+\nabla\bm{u}^T)) - (\bm{v},\bm{u}\cdot\nabla\bm{u}) , \label{eqn:wr_ns}\\
    (\nabla \cdot \bm{u}, q)&=0 . \label{eqn:wr_cont}
\end{align}
The $\mathbb{P}^P-\mathbb{P}^P$ velocity-pressure spaces are used \cite{deville2002high}.  The discretization up to this point is equivalent to a continuous finite-element discretization, but what separates SEM is the particular choice of basis and quadrature rule that will be discussed next.

Globally, each component of the velocity is represented on a Lagrange interpolating basis:
\begin{align}
    u(\bm{x}) = \sum_{i=1}^{N_c} \underline{u}_i \phi_i(\bm{x}) , \label{eqn:global_basis}
\end{align}
where $\underline{u}$ is the vector of basis coefficients, $\underline{u}_i$ is the $i$\textsuperscript{th} basis coefficient, $\phi_i(\bm{x})$ are continuous basis functions on $\Omega$, and $N_c$ is the total number of coefficients for all basis functions. This form is also used for the pressure. 

\begin{figure}[]
\centering
\includegraphics[width=0.5\textwidth,trim={0 0 0 0cm}]{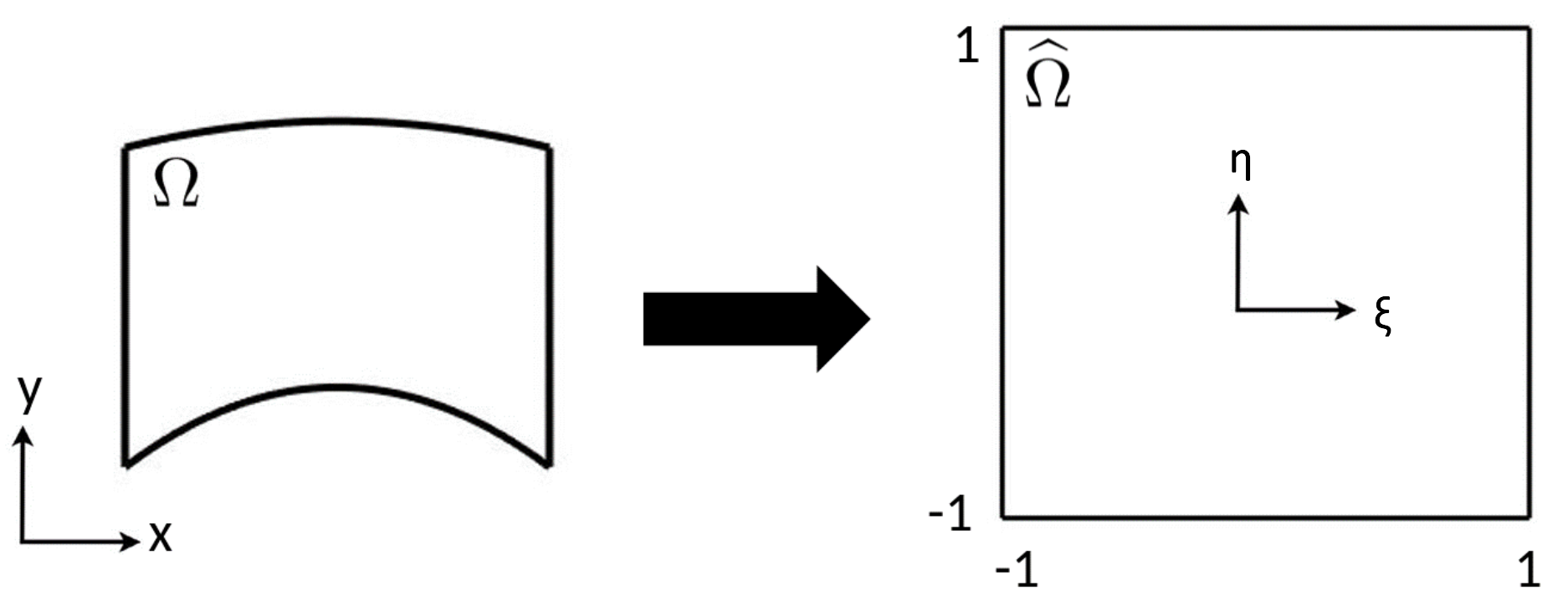}%
\caption{Schematic of mapping a physical element to a reference element.}
\label{fig:map_ref}
\end{figure}

In practice, the global form of the solution is never used. Instead, the domain, $\Omega$, is partitioned into $N_e$ non-overlapping discrete elements such that $\Omega = \cup_{e=1}^{N_e} \Omega_e$ and the solution is represented with local basis functions such that $\phi_i^e$ is non-zero in element $e$ and zero everywhere else. In this study, only tensor-product elements are used. Within each element the basis functions are formed by isoparametrically mapping the element to a reference element $\widehat{\Omega}\coloneqq [-1,1]^d$ with dimension, $d$, as schematically shown in Fig.~\ref{fig:map_ref} \cite{deville2002high}. From there, the solution is approximated by a high-order polynomial of order $P$:
\begin{align}
    u = \sum_{i=1}^{N_b} \underline{u}^e_i\phi^e_i(\bm{x}) = \sum_{i=1}^{P+1}\sum_{j=1}^{P+1}\sum_{k=1}^{P+1} \underline{u}^e_{ijk} h_i(\xi) h_j(\eta) h_k(\zeta), \label{eqn:local_basis}
\end{align}
where $N_b=(P+1)^d$ is the number of basis functions in an element, $d$ is the dimension of the problem, and $h_i$ is the $i$\textsuperscript{th} 1D basis function used in the tensor product. In general, the 1D basis functions could be any set of polynomial basis functions. 

For SEM, the 1D basis functions are chosen as the Lagrangian polynomial constructed on $P+1$ Gauss-Lobatto-Legendre (GLL) points in the reference space coordinates $\xi-\eta-\zeta$. A Lagrangian basis defined by a quadrature rule is chosen so that the solution can be numerically integrated with only nodal evaluations and will not require interpolation \cite{deville2002high}. The GLL points were chosen because they are the most accurate quadrature rule for polynomials that contains the end points of the element \cite{ascher2011first}. Having nodes on the element boundaries allows continuity between elements and boundary conditions to be enforced without interpolation. The basis is constructed with a tensor-product of 1D bases because the tensor-product structure can be exploited when computing derivatives and in other operations, which greatly reduces the number of floating point operations (FLOPS) for such operations \cite{deville2002high}.

In order to solve for the basis coefficients, Eqs.~\eqref{eqn:wr_ns}-\eqref{eqn:wr_cont} are formulated for each individual element by substituting the local solution representation from Eq.~\eqref{eqn:local_basis} for $\bm{u}$ and $p$. This equation is satisfied for each of the local basis functions $\phi^e_i(\bm{x})$ as the test function $\bm{v}$. By doing this, an element local system of equations is formed:
\begin{align}
\mathcal{M}^e \frac{d \underline{\bm{u}}^e}{dt} &=  \frac{1}{\rho}\mathcal{D}^{e,T} \underline{p}^e - \nu\mathcal{K}^e \underline{\bm{u}}^e - \mathcal{C}^e (\bm{u}^e), \\
    \mathcal{D}^e\underline{\bm{u}}^e &= 0, 
\end{align}
where $\mathcal{M}^e$ is the elemental mass matrix, $\mathcal{D}^e$ is the elemental differentiation matrix, $\mathcal{K}^e$ is the elemental stiffness matrix, $\mathcal{C}^e$ is the elemental non-linear convection operator, and $\underline{\bm{u}}^e$ is the elemental vector of coefficients $[\underline{u}_0, \underline{u}_1, ..., \underline{u}_{N_b}]$. The viscosity, $\nu$, is assumed to be constant.  The entries of these matrices are defined as
\begin{align}
    \mathcal{M}^e_{ij} &= \eleint{\phi_i^e \phi_j^e}, \qquad  \mathcal{D}^e_{ij} = \eleint{\frac{d \phi^e_j}{d x}\phi^e_i}, \label{eqn:mass_diff_mat} \\ 
    \mathcal{K}^e_{ij} &= \eleint{\nabla \phi_j^e \cdot \nabla \phi_i^e}, \qquad  \mathcal{C}^e_j = \eleint{\phi^e_j \bm{u}^e\cdot\nabla\bm{u}^e}. \label{eqn:stiff_conv_mat}
\end{align}

Each of the entries in the matrices, as shown in Eqs.~\eqref{eqn:mass_diff_mat}-\eqref{eqn:stiff_conv_mat}, are found by computing an integral over the element. These integrals are approximated using numerical quadrature. As previously mentioned, the numerical quadrature rule used in SEM is the same $P+1$ point GLL rule used to define the 1D basis functions in each direction. This quadrature rule exactly integrates polynomials of order $2P-1$. Hence, this rule is inexact for the mass matrix, because its entries are polynomials of order $2P$ \cite{maday1990optimal}. However, the impact of this inexactness is minimized for higher $P$ because the error from inexact integration is of the same order as the error of the polynomial expansion \cite{10.1093/acprof:oso/9780198528692.001.0001}. The advantage of this choice of quadrature rule and basis is that each basis function is only non-zero at a single quadrature point. As a result, the basis functions have discrete orthogonality and the mass matrix is a diagonal matrix and can be easily solved. Furthermore, evaluating the solution at each quadrature point is equal to the coefficient values at the nodes and no interpolation is needed. Hence, the choice of quadrature rule and basis functions enables efficient evaluations.

One complication during the computation of the matrix terms comes when computing $\mathcal{C}^e$ in Eq.~\eqref{eqn:stiff_conv_mat}. This term comes from the non-linear convection term in the Navier-Stokes equations.  Because of the non-linearity, if insufficient quadrature is used, high wavenumbers are not captured by the scheme and the energy that is supposed to be contained in those modes is aliased to lower wavenumbers, causing instability and inaccuracies in turbulent flow simulations \cite{kirby2003aliasing}. This issue can be rectified by using a higher-order quadrature rule when computing $\mathcal{C}^e$, which is called overintegration or dealiasing. Because of the order of the polynomials being integrated in the term, an overintegration rule of a $1.5(P+1)$ Gauss-Legendre (GL) points is sufficient to resolve the issue \cite{kirby2003aliasing}.

Once the local system of equations is formed for each element, they are combined to form a global system of equations. SEM requires the solution to have $C^0$ continuity between elements. To enforce this, co-located nodes are equated at each time step. These conditions are combined with the elemental systems of equations and setting nodal values on the boundary and combining equations with a direct stiffness sum to form a global system of equations for all of the local coefficients. This forms a global system of equations:
\begin{align}
\mathcal{M} \frac{d \underline{\bm{u}}}{dt} &=  \frac{1}{\rho}\mathcal{D}^{T} \underline{p} - \nu\mathcal{K} \underline{\bm{u}} - \mathcal{C} (\bm{u}), \\
    \mathcal{D}\underline{\bm{u}} &= 0,
\end{align}
where $\underline{\bm{u}}$ is the vector of global coefficients \cite{deville2002high}.

In summary, the solution procedure is to first compute the integrals in Eqs.~\eqref{eqn:wr_ns}-\eqref{eqn:wr_cont} using GLL quadrature to form a local system of equations for each element. Then a direct stiffness summation is used to couple neighboring elements and to create a global system of equations to solve for the solution coefficients \cite{fischer2017recent}. Finally, the equations are discretized in time and the global system of equations is solved for the basis coefficients at each time step.

The time-discretization chosen in this study is a semi-implicit BDF$k$/EXT$k$ scheme where the linear terms are treated implicitly with a $k$th-order backward difference formula and the non-linear terms are treated explicitly with a $k$th-order extrapolation \cite{fischer2017recent}.  The resulting discretization in time at step $t^n$ is 
\begin{align}
\sum_{j=0}^k\frac{\beta_j}{\Delta t}\mathcal{M} \underline{\bm{u}}^{n-j} &=  \frac{1}{\rho}\mathcal{D}^{T} \underline{p}^n - \nu\mathcal{K} \underline{\bm{u}}^n - \sum_{j=1}^{k}\alpha_{j}\mathcal{C} (\bm{u}^{n-j}), \label{eqn:wr_ns_dt}\\
    \mathcal{D}\underline{\bm{u}}^n &= 0, \label{eqn:wr_cont_dt}
\end{align}  
where $\beta_j$ and $\alpha_j$ are standard BDF$k$/EXT$k$ coefficients \cite{fischer2017recent}. The scheme is accurate to $O(\Delta t^k)$. Rearranging Eqs.~\eqref{eqn:wr_ns_dt} and \eqref{eqn:wr_cont_dt} so that all terms at $t^n$ are on the left side gives
\begin{align}
\frac{\beta_0}{\Delta t}\mathcal{M} \underline{\bm{u}}^{n} + \nu\mathcal{K} \underline{\bm{u}}^n - \frac{1}{\rho}\mathcal{D}^{T} \underline{p}^n &= -\sum_{j=1}^{k} (\alpha_{j}\mathcal{C} (\bm{u}^{n-j})+\frac{\beta_j}{\Delta t}\mathcal{M} \underline{\bm{u}}^{n-j})  \label{eqn:wr_ns_dt2}\\
    \mathcal{D}\underline{\bm{u}}^n &= 0. \label{eqn:wr_cont_dt2}
\end{align}  
Values of $k=1,2,3$ can be used. The implicit part of the scheme extends the stability of the scheme, while treating the non-linear term explicitly avoids having to solve a non-linear system of equations at each time step \cite{patera1984spectral}. In order to reach the final solution time as quickly as possible, the timestep was adapted throughout the simulation to meet a set Courant-Friedrichs-Lewy (CFL) constraint. For notational convenience, the discrete Helmholtz matrix is defined as $\mathcal{H} =  \frac{\beta_0}{\Delta t}\mathcal{M} + \nu\mathcal{K}$ making the overall system of equations
\begin{align}
\mathcal{H} \underline{\bm{u}}^{n} - \frac{1}{\rho}\mathcal{D}^{T} \underline{p}^n &= -\sum_{j=1}^{k} (\alpha_{j}\mathcal{C} (\bm{u}^{n-j})+\frac{\beta_j}{\Delta t}\mathcal{M} \underline{\bm{u}}^{n-j})  \label{eqn:wr_ns_dt3}\\
    \mathcal{D}\underline{\bm{u}}^n &= 0. \label{eqn:wr_cont_dt3}
\end{align}  
Additionally, to improve the conditioning of the system, the change in the coefficients $\Delta \underline{\bm{u}}^{n} = \underline{\bm{u}}^{n} - \underline{\bm{u}}^{n-1}$ is solved for:
\begin{align}
\mathcal{H} \Delta \underline{\bm{u}}^{n} - \frac{1}{\rho}\mathcal{D}^{T} \underline{p}^n &= -\mathcal{H} \Delta \underline{\bm{u}}^{n-1}-\sum_{j=1}^{k} (\alpha_{j}\mathcal{C} (\bm{u}^{n-j})+\frac{\beta_j}{\Delta t}\mathcal{M} \underline{\bm{u}}^{n-j})  \label{eqn:wr_ns_dt4}\\
   \mathcal{D}\Delta\underline{\bm{u}}^n &= -\mathcal{D}\underline{\bm{u}}^{n-1}, \label{eqn:wr_cont_dt4} \\
   \underline{\bm{u}}^n &= \underline{\bm{u}}^{n-1} + \Delta\underline{\bm{u}}^n .
\end{align}  
Assuming $\bm{u}^{n-1}$ is $C^0$ continuous, $\bm{u}^{n}$, will be $C^0$ continuous if $\Delta\bm{u}^{n}$ is also $C^0$ continuous. Hence, because the operators in Eq.~\eqref{eqn:wr_ns_dt4}-\eqref{eqn:wr_cont_dt4} enforce $C^0$ continuity of $\Delta\bm{u}^{n}$, $\bm{u}^{n}$ will maintain the appropriate continuity at all timesteps.

\section{Enrichment Method}
\label{sec:enrich}
\begin{figure}[]
\centering
\includegraphics[width=0.7\textwidth,trim={0 0 0 0cm}]{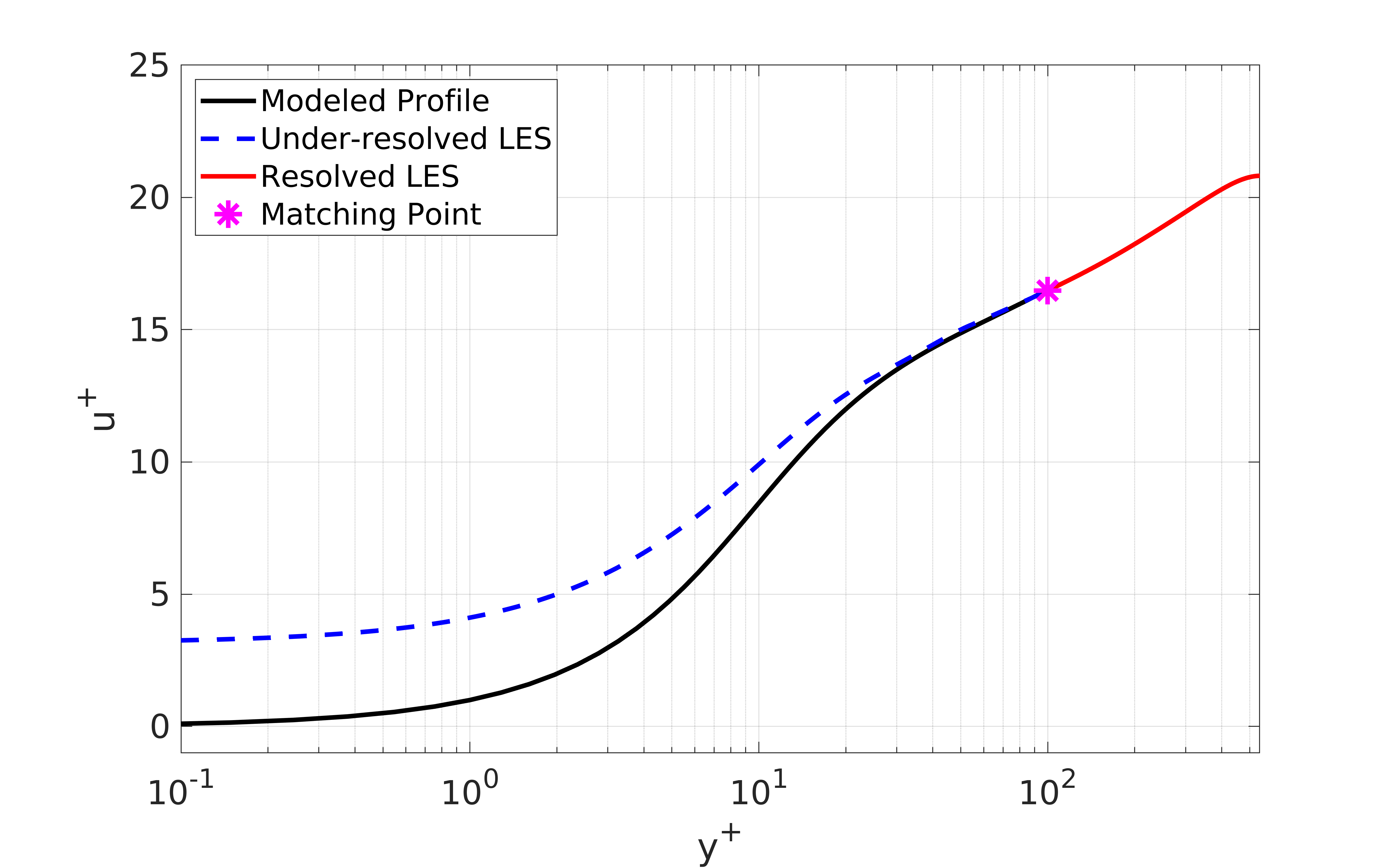}%
\caption{Schematic of the shear stress wall-model showing the under-resolved region near the wall and the resolved region connected by the matching point.}
\label{fig:sswm_schematic}
\end{figure}

The most common WMLES methods sample the solution away from the wall in the fully resolved region and apply a modeled shear stress. The modeled shear stress is applied as a slip-wall boundary condition. The solution near the wall is under resolved and serves only to pass the shear stress to the bulk. This section of the domain can contain oscillations and large amounts of slip, resulting in an unphysical section of the domain as shown in Fig.~\ref{fig:sswm_schematic}. The unphysical section in the domain can cause inaccuracies when computing wall quantities and coupling physics in the near-wall region, such as Lagrangian particles which are very sensitive to the flow and geometry near the wall \cite{CHING2020109096}. To address these issues, we developed a solution enrichment based wall-model. Instead of modifying the boundary conditions at the wall, we modify the solution representation to capture the modeled boundary layer profile without unphysical oscillations due to under-resolution.

The overall idea is that in a subset of the domain, $\Omega_{\psi}$, the polynomial solution is enriched with an enrichment function $\bm{\psi}(\bm{x},t)$. This function represents a feature in the solution that cannot be accurately captured by the polynomial basis. Typically such features are non-polynomial functions with too large of gradients to capture with the polynomial basis without unphysical oscillations. For this application, the enrichment function will capture the mean velocity profile of the turbulent boundary layer and will be discussed in detail in Section~\ref{sec:en_fun}. The enriched solution representation is: 
\begin{align}
    \bm{u}(\bm{x},t) = \upoly(\bm{x},t) + \bm{\psi}(\bm{x},t), \label{eqn:u_en}
\end{align}
where $\upoly$ is the polynomial component of the solution defined on the standard SEM basis:
\begin{align}
    \upoly = \sum_i^{N_b} \underline{\bm{u}}_i\phi_i(\bm{x}) .
\end{align}
With this representation, the enrichment function is able to capture the difficult flow feature and the polynomials can capture other features or fluctuations. 

The enriched SEM begins by putting the governing equations into weighted residual form as shown in Eqs.~\eqref{eqn:wr_ns1}-\eqref{eqn:wr_cont1}. However, for the enriched SEM, the solution is $(\bm{u},p)\in X_{b,\psi}^P(\Omega)\times Y^P(\Omega)$, where $X_{\psi}\subset H^1(\Omega)$ such that $X_\psi^P = \{ u \,|\, u = u_p + \psi, u_p \in X^P(\Omega_\psi) \cap X^P(\Omega_p), \psi \in H^1(\Omega_{\psi}) \cap H^1(\Omega_{p}), \psi(\bm{x})=0 \, \forall \bm{x} \in \Omega_p\}$, where $\Omega_\psi$ and $\Omega_p$ are the enriched and unenriched parts of the domain respectively, such that $\Omega_\psi \cup \Omega_p = \Omega$ and $\Omega_\psi$ and $\Omega_p$ do not overlap. $X_{b,\psi}$ is the subset of $X_{\psi}$ such that Dirichlet boundary conditions are satisfied on the boundary $\partial \Omega$. As is done in the traditional SEM, the weighted residual form is integrated by parts to give Eqs.~\eqref{eqn:wr_ns}-\eqref{eqn:wr_cont}.
From here, the enriched solution representation from Eq.~\eqref{eqn:u_en} is substituted into the weighted residual form:
\begin{align}
    \frac{\partial}{\partial t} (\bm{v},\upoly+\bm{\psi}) =& (\nabla \bm{v}, p) - (\nabla\bm{v}, \left[\nu\left(  \nabla(\upoly+\bm{\psi})+\nabla(\upoly+\bm{\psi})^T\right)\right]) \label{eqn:en_ns_weak_ibp}\\
    &- (\bm{v},(\upoly+\bm{\psi})\cdot \nabla (\upoly+\bm{\psi})) ,\nonumber \\
(\nabla \cdot (\upoly+\bm{\psi}),q) =& \,  0 .
\end{align}
Then equations are algebraically manipulated to separate the enrichment terms from the polynomial terms:
\begin{align}
    \frac{\partial}{\partial t}(\bm{v},\upoly) &= (\nabla \cdot \bm{v}, p) - (\nabla\bm{v},\nu(\nabla\upoly+\nabla\upoly^T)) - (\bm{v},\upoly\cdot\nabla\upoly)  \label{eqn:en_wr_ns}\\
    &- (\nabla\bm{v}, \nu\left( \nabla\bm{\psi}+\nabla\bm{\psi}^T\right)) -(\bm{v},\upoly\cdot\nabla\bpsi) -(\bm{v},\bpsi\cdot\nabla\upoly) \nonumber \\
   &-(\bm{v},\bpsi\cdot\nabla\bpsi) - \frac{d}{dt}(\bm{v},\bm{\psi}), \nonumber \\
    (\nabla \cdot \upoly, q) +& (\nabla \cdot \bpsi, q)=0  .\label{eqn:en_wr_cont}
\end{align}
For the application to WMLES, we assume the enrichment function is constant within a timestep and is adapted explicitly using matching points, which will be detailed in Sec.~\ref{sec:en_fun}. Hence, the time derivative of the enrichment function is put to the right hand side of Eq.~\eqref{eqn:en_wr_ns}. With the governing equations in this form, the left-hand side and the first three terms on the right-hand side are the same as in the traditional SEM and the last five terms on the right-hand side are additions from the enrichment \cite{brill2022enrichment}. This separability allows the presented enrichment method can be easily implemented with the addition of associated terms into existing solvers as right-hand side terms.

To fully define these terms, the local systems of equations are formed by substituting in the local solution representation and letting the test function be the local basis functions, $\phi_i^e(\bm{x})$. Through this the local system of equations is formed:
\begin{align}
\mathcal{M}^e \frac{d \underline{\bm{u}}^e_p}{dt} &=  \frac{1}{\rho}\mathcal{D}^{e,T} \underline{p}^e - \nu\mathcal{K}^e \underline{\bm{u}}^e_p - \mathcal{C}^e (\bm{u}^e_p) - \mathcal{V}^e_\psi -\mathcal{C}^e_\psi(\bm{u}^e_p) - \frac{d}{dt}(\mathcal{M}^e_\psi), \label{eqn:en_mat_form} \\
    \mathcal{D}^e\underline{\bm{u}}^e_p + \mathcal{D}^e_\psi &= 0,  \label{eqn:en_mat_form_P}
\end{align}
where $\mathcal{V}_\psi$ is the viscous enrichment operator, $\mathcal{C}_\psi$ is the convective enrichment operator, $\mathcal{D}_\psi$ is differentiation enrichment operator, and $\mathcal{M}_\psi$ is the enrichment mass operator. The entries of these matrices are defined as
\begin{align}
    \mathcal{V}^e_{\psi,j} &= \eleint{\nu \nabla\phi_j^e \cdot (\nabla \bm{\psi}^e+\nabla \bm{\psi}^{e,T})}, \qquad  \mathcal{D}^e_{\psi,j} = \eleint{\nabla \cdot \bm{\psi}^e \phi^e_j}, \label{eqn:en_V_D_mat} \\ 
    \mathcal{C}^e_{\psi,j}(\bm{u}^e_p) &= \eleint{\phi_j^e (\bm{u}^e_p\cdot \nabla \bm{\psi}^e + \bm{\psi}^e\cdot\nabla \bm{u}^e_p + \bm{\psi}^e\cdot\nabla\bm{\psi}^e)} , \label{eqn:en_C_mat} \\
    \mathcal{M}^e_{\psi,j} &= \eleint{\phi_j^e \bm{\psi}^e}. \label{eqn:en_M_mat}
\end{align}
As with the standard SEM matrix entries, these are computed with numerical quadrature over the elements. However, the enrichment function contains a feature that the polynomial basis cannot resolve. Hence, quadrature rules designed for order $P$ polynomials, like the standard $P+1$ point GLL quadrature rule, are likely to be inaccurate when integrating the enrichment function. To address this issue, a quadrature rule designed for the enrichment function is used. If it can be integrated exactly or accurately with a specific quadrature rule, that should be used. If the enrichment function's structure cannot be exploited, we have found using a higher-order GL or GLL quadrature rule to be sufficient for smooth enrichment functions. Numerical accuracy tests are performed to determine a heuristic for a particular enrichment function and basis order.

Now we will discuss the time discretization applied to the local systems of equations. As before, Eqs.~\eqref{eqn:en_mat_form}-\eqref{eqn:en_mat_form_P} are discretized in time using the BDF$k$/EXT$k$ scheme where the viscous terms are treated implicitly and the convective terms are treated explicitly:
\begin{align}
\sum_{j=0}^k\frac{\beta_j}{\Delta t}\mathcal{M}^e \underline{\bm{u}}^{e,n-j}_p =&  \frac{1}{\rho}\mathcal{D}^{e,T} \underline{p}^{e,n} - \nu\mathcal{K}^e \underline{\bm{u}}^{e,n}_p - \sum_{j=1}^{k}\alpha_{j}\mathcal{C}^e (\bm{u}^{e,n-j}_p) \label{eqn:en_wr_ns_dt} \\
&- \mathcal{V}_\psi^{e,n} -\sum_{j=1}^k \alpha_j\mathcal{C}^{e,n-j}_\psi(\bm{u}_p^{e,n-j}) - \sum_{j=0}^k\frac{\beta_j}{\Delta t}\mathcal{M}_\psi^{e,n-j}, \nonumber\\
    \mathcal{D}^e\underline{\bm{u}}^{e,n}_p  + \mathcal{D}^e_\psi =& 0. \label{eqn:en_wr_cont_dt}
\end{align}  
Rearranging the equations to group the variables being solved for gives:
\begin{align}
\mathcal{H}^e\underline{\bm{u}}^{e,n}_p - \frac{1}{\rho}\mathcal{D}^{e,T} \underline{p}^{e,n} &= - \sum_{j=1}^k\left(\alpha_j\mathcal{C}^e(\bm{u}^{e,n-j}_p)+\frac{\beta_j}{\Delta t}\mathcal{M}^e\underline{\bm{u}}^{e,n-j}_p\right) \label{eqn:en_wr_ns_dt_H} \\ 
& - \sum_{j=1}^k\left(\alpha_j\mathcal{C}_\psi^{e,n-j}(\bm{u}^{e,n-j}_p)+\frac{\beta_j}{\Delta t}\mathcal{M}_\psi^{e,n-j}\right)  
-\left(\frac{\beta_0}{\Delta t}\mathcal{M}^{e,n}_\psi+\mathcal{V}^{e,n}_\psi\right) \nonumber \\
    \mathcal{D}\underline{\bm{u}}^n_p  =& - \mathcal{D}_\psi . \label{eqn:en_wr_cont_dt_H}
\end{align}  
In matrix form, as well as variational form, the enrichment terms can be separated from the standard SEM terms. Finally, the equations are modified to solve for the change in the coefficients:
\begin{align}
\mathcal{H}^e\Delta\underline{\bm{u}}^{e,n}_p - \frac{1}{\rho}\mathcal{D}^{e,T} \Delta\underline{p}^{e,n} &= -\mathcal{H}^e\Delta\underline{\bm{u}}^{e,n-1}_p + \frac{1}{\rho}\mathcal{D}^{e,T} \Delta\underline{p}^{e,n-1} \label{eqn:en_wr_ns_dt_H_du_local}\\ 
& - \sum_{j=1}^k\left(\alpha_j\mathcal{C}^e(\bm{u}^{e,n-j}_p)+\frac{\beta_j}{\Delta t}\mathcal{M}^e\underline{\bm{u}}^{e,n-j}_p\right)  \nonumber \\ &- \sum_{j=1}^k\left(\alpha_j\mathcal{C}_\psi^{e,n-j}(\bm{u}^{e,n-j}_p)+\frac{\beta_j}{\Delta t}\mathcal{M}_\psi^{e,n-j}\right) 
-\left(\frac{\beta_0}{\Delta t}\mathcal{M}^{e,n}_\psi+\mathcal{V}^{e,n}_\psi\right) \nonumber \\
    \mathcal{D}^e\Delta\underline{\bm{u}}^{e,n}_p   =& - \mathcal{D}^e_\psi -\mathcal{D}^e\underline{\bm{u}}^{e,n-1}_p, \label{eqn:en_wr_cont_dt_H_du_local} \\
    \underline{\bm{u}}^{e,n}_p &= \underline{\bm{u}}^{e,n-1}_p + \Delta\underline{\bm{u}}^{e,n}_p .
\end{align}  
Because the enrichment function is constant within each timestep, the change in the enriched solution is equal to the change in the polynomial component: $\Delta\bm{u}^{n} = \Delta\bm{u}^{n}_p$. Further, Lemma~\ref{lem:cont} proves that $\Delta u_p^n \in H^1(\Omega)$. Hence, the standard SEM scatter-gather operations ensure that the solution is continuous \cite{fischer2017recent}:
\begin{align}
    \mathcal{H}\Delta\underline{\bm{u}}^{n}_p - \frac{1}{\rho}\mathcal{D}^{T} \Delta\underline{p}^{n} &= -\mathcal{H}\Delta\underline{\bm{u}}^{n-1}_p + \frac{1}{\rho}\mathcal{D}^{T} \Delta\underline{p}^{n-1} \label{eqn:en_wr_ns_dt_H_du}\\ 
& - \sum_{j=1}^k\left(\alpha_j\mathcal{C}(\bm{u}^{n-j}_p)+\frac{\beta_j}{\Delta t}\mathcal{M}\underline{\bm{u}}^{n-j}_p\right)  \nonumber \\ &- \sum_{j=1}^k\left(\alpha_j\mathcal{C}_\psi^{n-j}(\bm{u}^{n-j}_p)+\frac{\beta_j}{\Delta t}\mathcal{M}_\psi^{n-j}\right) 
-\left(\frac{\beta_0}{\Delta t}\mathcal{M}^{n}_\psi+\mathcal{V}^{n}_\psi\right) \nonumber \\
    \mathcal{D}\Delta\underline{\bm{u}}^{n}_p   =& - \mathcal{D}_\psi -\mathcal{D}\underline{\bm{u}}^{n-1}_p.\label{eqn:en_wr_cont_dt_H_du} 
\end{align}

\begin{lemma}
\label{lem:cont}
If $\psi^n$ is continuous within a timestep, the change in the solution, $\Delta u_p^n$ is in $X^P(\Omega)$.
\end{lemma}
\begin{proof}
    Let $u_{p,\psi}$ and $u_{p,p}$ be the polynomial portions of the solution in $\Omega_\psi$ and $\Omega_p$ respectively. $u_{p,\psi}^n$ and $u_{p,\psi}^{n-1}$  are in $X^P(\Omega_\psi)$. Hence $\Delta u_{p,\psi}^n \in X^P(\Omega_\psi)$. Similarly, $\Delta u_{p,p}^n \in X^P(\Omega_p)$. Let $\Gamma_{\psi,p}$ be the boundary between $\Omega_\psi$ and $\Omega_p$. Within $\Omega_{\psi}$, the value of $u(\Gamma_{\psi,p}) = u_{p,\psi}(\Gamma_{\psi,p}) + \psi(\Gamma_{\psi,p})$. Hence, because $\psi^n$ is constant within the timestep,
    \begin{align}
    \Delta u^n(\Gamma_{\psi,p}) &= u^n_{p,\psi}(\Gamma_{\psi,p}) + \psi^n(\Gamma_{\psi,p}) - u^{n-1}_{p,\psi}(\Gamma_{\psi,p}) - \psi^n(\Gamma_{\psi,p}) ,\\
    &= u^n_{p,\psi}(\Gamma_{\psi,p})  - u^{n-1}_{p,\psi}(\Gamma_{\psi,p}) ,\\
    &= \Delta u^n_{p,\psi}.
    \end{align}
    Similarly, within $\Omega_p$, $u(\Gamma_{\psi,p}) = u_{p,p}(\Gamma_{\psi,p})$, so 
    \begin{align}
        \Delta u^n(\Gamma_{\psi,p}) &= u^n_{p,p}(\Gamma_{\psi,p})- u^{n-1}_{p,p}(\Gamma_{\psi,p}), \\
        &= \Delta u^n_{p,p}.
    \end{align} 
    Because $u^n, u^{n-1} \in H^1(\Omega)$, $\Delta u^n(\Gamma_{\psi,p})$ is equal between both parts of the domain. Hence $\Delta u^n_{p,\psi}(\Gamma_{\psi,p}) = \Delta u^n_{p,p}(\Gamma_{\psi,p})$. $\Delta u_p^n$ is in $X^P(\Omega_\psi)\cap X^P(\Omega_p)$ and is continuous at $\Omega_\psi \cap \Omega_p$. Hence $\Delta u_p^n \in X^P(\Omega)$.
\end{proof}

The solution procedure for the polynomial coefficients is exactly the same as the standard SEM and high-order polynomial portion of the solution representation is maintained. Hence, the addition of the enrichment function improves the method's ability to capture the targeted flow feature without affecting the high-order accuracy.

\section{Enrichment Function}
\label{sec:en_fun}

Shear stress wall-models modify the no-slip boundary condition so that the solution representation can match the modeled gradient with large elements in the near-wall region without oscillations. The enrichment wall-model uses the enrichment framework described in Sec.~\ref{sec:enrich} to modify the solution representation to capture the large gradients in the boundary layer without modifying the boundary conditions. To achieve this, the enrichment function is designed to capture the mean velocity profile of the turbulent boundary layer, which leaves the polynomial part of the solution to represent the turbulent fluctuations.

The enrichment function is constructed such that it captures large gradients of the law-of-the-wall in the wall-normal direction and remains continuous in the wall-parallel directions. The goal of WMLES is to use large elements to represent the boundary layer. Hence, for this study, the enriched domain, $\Omega_\psi$, is defined as only the wall-adjacent elements. To develop the wall-model enrichment function, we consider a wall-adjacent element in reference space and without loss of generality, assume $\eta$ is the wall-normal reference direction. The nodes on the wall are indexed form $1$ to $P+1$ with $i$ and $j$ for the $\xi$ and $\zeta$ reference spatial directions, respectively. At each node on the wall, a law-of-the-wall function is defined as a function of the wall-normal direction, $\bm{\psi}_{w,(i,j)}(\eta)$ to represent the boundary layer profile. An SEM polynomial basis is used in the wall-parallel directions to ensure continuity. The resulting enrichment function in element $e$ is
\begin{align}
    \bm{\psi}_w^e(\bm{\xi}) =& \sum_{i=1}^{P+1} \sum_{j=1}^{P+1} \bm{\psi}^e_{w,(i,j)}(\eta) \phi^e_i(\xi) \phi^e_j(\zeta) , \label{eqn:gen_enrich_fnc} 
\end{align}
where $\bm{\xi}$ is the vector of reference coordinates and $\phi_i$ is the $i$th 1D nodal GLL basis function.

The law-of-the-wall functions are defined at each node so they can be locally adapted to the flow features. This is done using matching points, like in shear stress wall-models. For each node on the wall $(i,j)$, the matching point is taken to be the corresponding node on the opposite end of the wall-adjacent element in the $\eta$ direction, because it is assumed that the entire law-of-the-wall profile is contained in the first element. In other studies, this choice has been shown to prevent log-layer mismatch \cite{frere2017application}. The wall-parallel velocity at the matching point and its distance from the wall in physical space are notated $u_{\|,(i,j)}$ and $h_{w,(i,j)}$, respectively. Then, an algebraic law-of-the-wall function $f_w(h_w,u_\tau)$ is chosen to model the mean velocity profile. Any algebraic law-of-the-wall function can be used. In this study, we use the Reichardt law-of-the-wall due to its popularity in WMLES methods, smoothness, and ease of computation:
\begin{align}
    f_w(h_w,u_{\tau}) = u_\tau\left(\frac{1}{0.41}\log(1+0.41y^+)+7.8(1-e^{-\frac{y^+}{11}}-\frac{y^+}{11}e^{-\frac{y^+}{3}})\right),
\end{align}
where $y^+=h_w u_{\tau}/\nu$ is the non-dimensional distance from the wall. The Spalding law-of-the-wall was also tested, but showed little effect on the solution and is computationally more expensive to evaluate \cite{spalding1961single}.
The law-of-the-wall function is fit to the resolved LES region by solving for the friction velocity that makes the modeled velocity equal to the the targeted LES velocity at the matching point, $u^*_{\|,(i,j)}$. The target LES velocity can be either the instantaneous or a time-averaged velocity \cite{larsson2016large}. In this study, it was seen that time-averaged velocities over a flow through time tended to be more stable than instantaneous values, so they will be used going forward. Using the  $u^*_{\|,(i,j)}$, $h_{w,(i,j)}$, and $f_w$, an implicit nonlinear equation is solved for the friction velocity at the node, $u_{\tau,(i,j)}$:
\begin{align}
    u^*_{\|,(i,j)} = f_w(h_{w,(i,j)},u_{\tau,(i,j)}) . \label{eqn:matching_gen_ij}
\end{align}
 Using the modeled friction velocity, the local law-of-the-wall function in the enrichment function is defined by evaluating the same algebraic wall function:
\begin{align}
    \bm{\psi}_{w,(i,j)}(\eta)=f_w(h_w(\eta),u_{\tau,(i,j)})\hat{\bm{t}}_{w,(i,j)}, \label{eqn:enrich_fnc_ij}
\end{align}
where $h_w(\eta)$ is the physical distance from the wall at point $\eta$ and $\hat{\bm{t}}_{w,(i,j)}$ is the unit tangent vector to the wall at node $(i,j)$ \cite{brill2022enrichment}. Hence the enrichment function in Eq.~\eqref{eqn:gen_enrich_fnc} is fully defined with the law-of-the-wall function capturing the large wall-normal gradients and polynomials keeping the function continuous in the wall-parallel directions.

The enrichment function is evolved in time to match local flow features. This can be achieved by recomputing the enrichment function explicitly at each timestep or at a set frequency. After the polynomial solution is computed at a timestep using Eqs.~\eqref{eqn:en_wr_ns_dt_H}-\eqref{eqn:en_wr_cont_dt_H}, a new enrichment function can be computed as described earlier. Then an $L_2$ projection is used to compute the new polynomial coefficients. From experimentation, it was seen that recomputing the enrichment function at every flow-through time was effective. Additionally, $\mathcal{M}_\psi$, $\mathcal{D}_\psi$, and $\mathcal{V}_\psi$ are only computed when the enrichment function is recomputed.

In order to accurately compute the integrals in Eqs.~\eqref{eqn:en_V_D_mat}-\eqref{eqn:en_M_mat} a quadrature must be tailored for the enrichment function. The enrichment wall-model is a continuous and smooth function with large gradients in only the wall-normal direction. Hence, overintegration in the wall-normal direction is used. From testing, we determined that overintegrating using $1.5(P+1)$ GL quadrature points in the $\eta$ direction and standard $P+1$ GLL quadrature points in the $\xi$ and $\zeta$ directions was sufficient to accurately resolve the integrals for a range of polynomial orders and Reynolds numbers. If the wall-model enrichment function or its derivative were discontinuous, a quadrature rule that could accurately integrate around the discontinuity, like a segmented quadrature rule, would be needed \cite{brill2020enriched}.

The resulting algorithm to compute the enrichment function is detailed in Algorithm~\ref{alg:comp_en}. This wall-model takes the matching point sufficiently far from the wall to prevent log-layer mismatch while the enrichment formulation ensures that the large gradients are captured in the solution without unphysical oscillations or unphysical boundary conditions.

\begin{algorithm}
\caption{Computing Enrichment Algorithm}\label{alg:comp_en}
\begin{algorithmic}
\For{each element on the wall}
\For{each node $(i,j)$ on the wall}
    \State Sample $u_{\|,(i,j)}$ and $h_{w,(i,j)}$ on the opposite end of the element.
    \State Solve for $u_{\tau,(i,j)}$ with Eq.~\eqref{eqn:matching_gen_ij}.
    \State Compute $\bm{\psi}_{w,(i,j)}$ with Eq.~\eqref{eqn:enrich_fnc_ij}.
\EndFor
\State Compute $\bm{\psi}_w^e$ with Eq.~\eqref{eqn:gen_enrich_fnc}.
\State Project the solution onto new enrichment function
\EndFor
\end{algorithmic}
\end{algorithm}

\section{Numerical Results}
\label{sec:num}
The test case considered for demonstration studies in the present work is the turbulent channel flow. The case is set up in non-dimensional form by choosing a domain of size $[2\pi h, 2h, \pi h]$ where $h$ is the half-height of the channel. The domain size ensures flow decorrelation of the turbulent structures. Periodic boundary conditions are used in the $x$ and $z$ directions. No-slip wall boundary conditions are used in the $y$ direction. In order to ensure that the flow in the channel is maintained, a variable forcing function that ensures a constant mass flow throughout the domain is applied \cite{scalo2015compressible}. The mass flow rate and domain are fixed, so only $\nu$ is changed to set $Re_\tau$. 

In order for the turbulent flow to become fully developed and statistically stationary, the flow is initialized with perturbations to speed the transition to turbulence. The initial conditions used in this work are
\begin{align}
u =& Re_\tau\left( \frac{1}{k}\log{(1+k y^+)} + (C - \frac{1}{k}\log{k})(1-e^{-\frac{y^+}{11}}-\frac{y^+}{11}e^{-\frac{y^+}{3}}\right) + \epsilon\beta \sin(\alpha x)\cos(\beta z) , \\
v =& \epsilon \sin(\alpha x)\cos(\beta z) , \\
w =& -\epsilon\alpha \cos(\alpha x)\sin(\beta z) ,
\end{align}
where $Re_\tau$ is the target friction Reynolds number, $C=5.17$ and $k=0.41$ are constants from the Reichardt law-of-the-wall, $\epsilon$ is the magnitude of the perturbation, and $\alpha$ and $\beta$ are the frequencies of the perturbations. For this study, values of $\epsilon=10^{-2}$, $\alpha = 23$, and $\beta = 13$ were used, so that turbulence tripped in the first flow-through time. Additionally, with a channel half-height of $h=1$, a constant mass flow rate of $\dot{m}=1$ was used. The Reynolds numbers targeted in this work are $Re_\tau = 543, 1000, 2000$ which have corresponding viscosities of $\nu = 10^{-4}, 5\times 10^{-5}, 2.3\times 10^{-5}$ and these conditions match those used by Lee et al. \cite{moser2015}.

The simulation is allowed to develop for 8 flow-through times and then statistics are computed for 300 flow-through times. The wall-model is turned on after 8 flow-through times and thereafter the flow is allowed to fully develop. For the enriched wall-model, the wall-parallel velocity used at the matching point is chosen to be the local time-averaged value over the last 50 time units, which corresponds to approximately 8 flow-through times. The Reichardt law-of-the-wall is used as the wall function in the enrichment function. The meshes used have $N$ uniform elements in each direction to ensure that they remain coarse in the near-wall region.

\subsection{Grid Resolution Study}
\label{sec:converge}
The enriched wall-model was tested on a number of coarse mesh resolutions, polynomial orders, and Reynolds numbers. These cases will show the generalizability of the wall-model and test its performance for challenging high Reynolds number conditions with larger gradients captured by the enrichment function. 

\begin{figure}[]
\centering
\includegraphics[width=\meansize\textwidth,trim={0 0 0 0cm}]{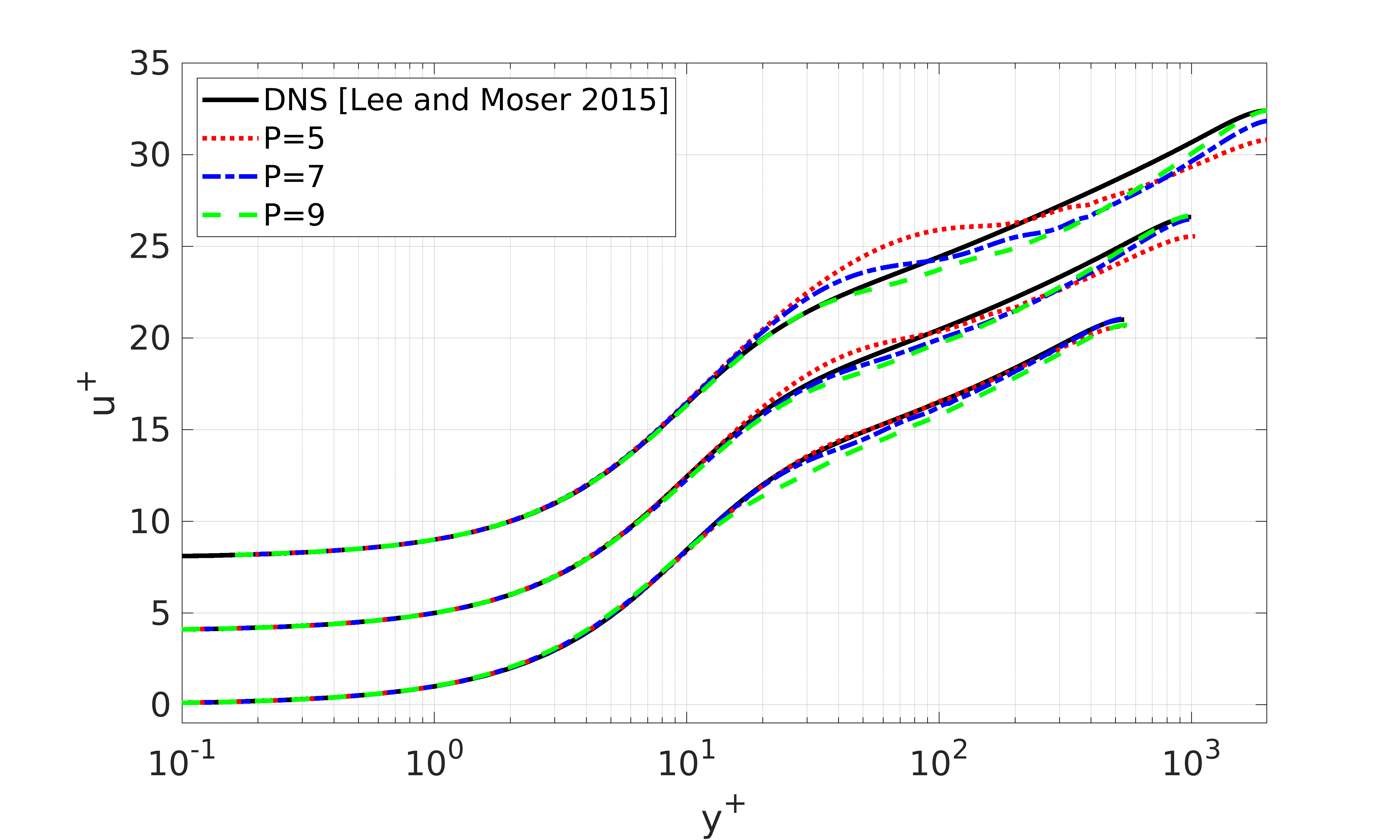}%
\caption{Mean velocity from WMLES of turbulent channel flow with $P=5,7,9$ order basis polynomials. Results for $Re_\tau=543$, $Re_\tau=1000$, and $Re_\tau=2000$, with $N=10$ are shown from bottom to top, shifted up by 4 for visibility. The solutions use a law-of-the-wall enrichment function in the wall-adjacent elements as the wall-model. The WMLES results are compared with the DNS data from~\cite{moser2015}. }
\label{fig:mean_conv10}
\end{figure}

\begin{figure}[]
\centering
\includegraphics[width=\meansize\textwidth,trim={0 0 0 0cm}]{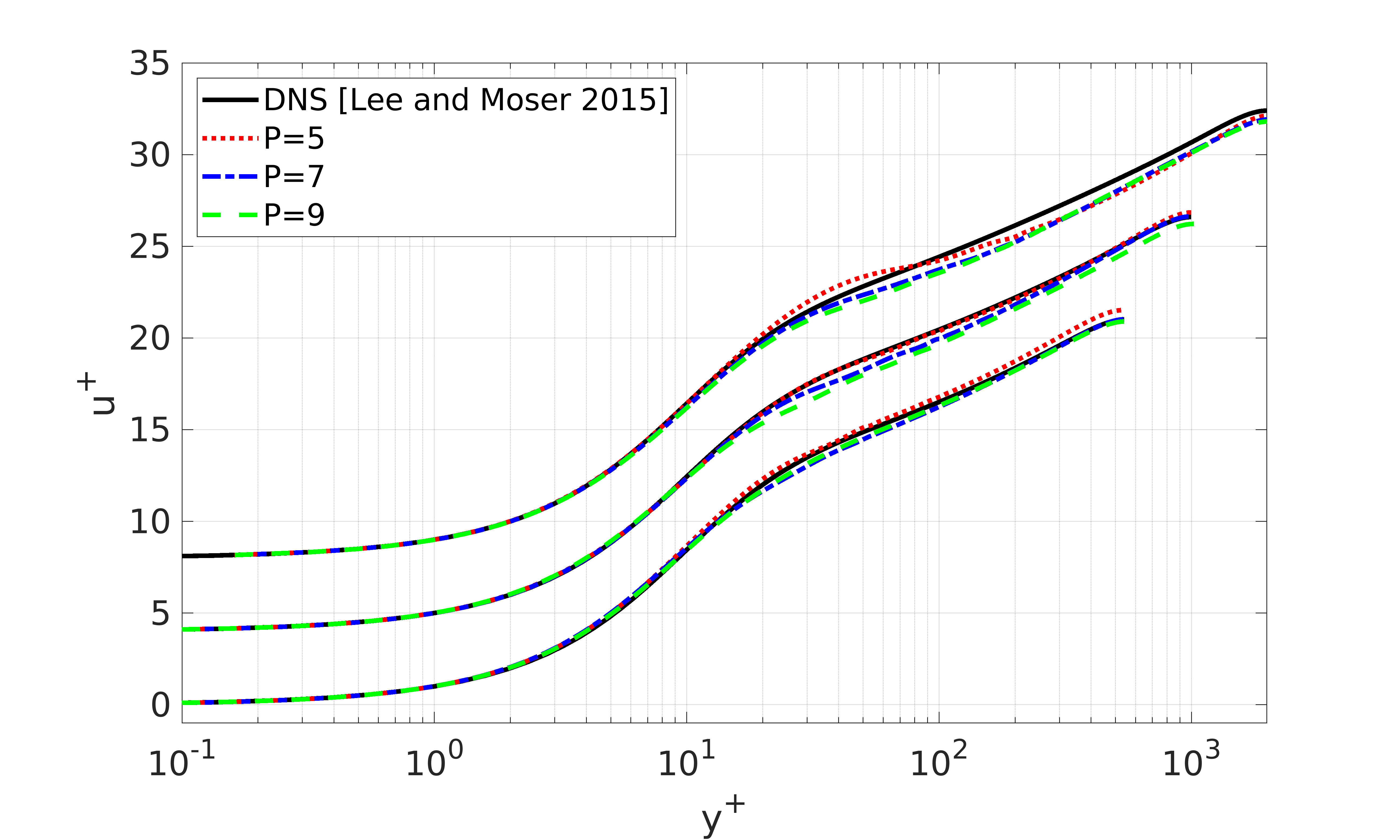}%
\caption{Mean velocity from WMLES of turbulent channel flow with $P=5,7,9$ order basis polynomials. Results for $Re_\tau=543$, $Re_\tau=1000$, and $Re_\tau=2000$, with $N=20$ are shown from bottom to top, shifted up by 4 for visibility. The solutions use a law-of-the-wall enrichment function in the wall-adjacent elements as the wall-model. The WMLES results are compared with the DNS data from~\cite{moser2015}. }
\label{fig:mean_conv20}
\end{figure}

\begin{figure}
     \centering
     \begin{subfigure}[b]{\rmssize\textwidth}
         \centering
         \includegraphics[width=\textwidth]{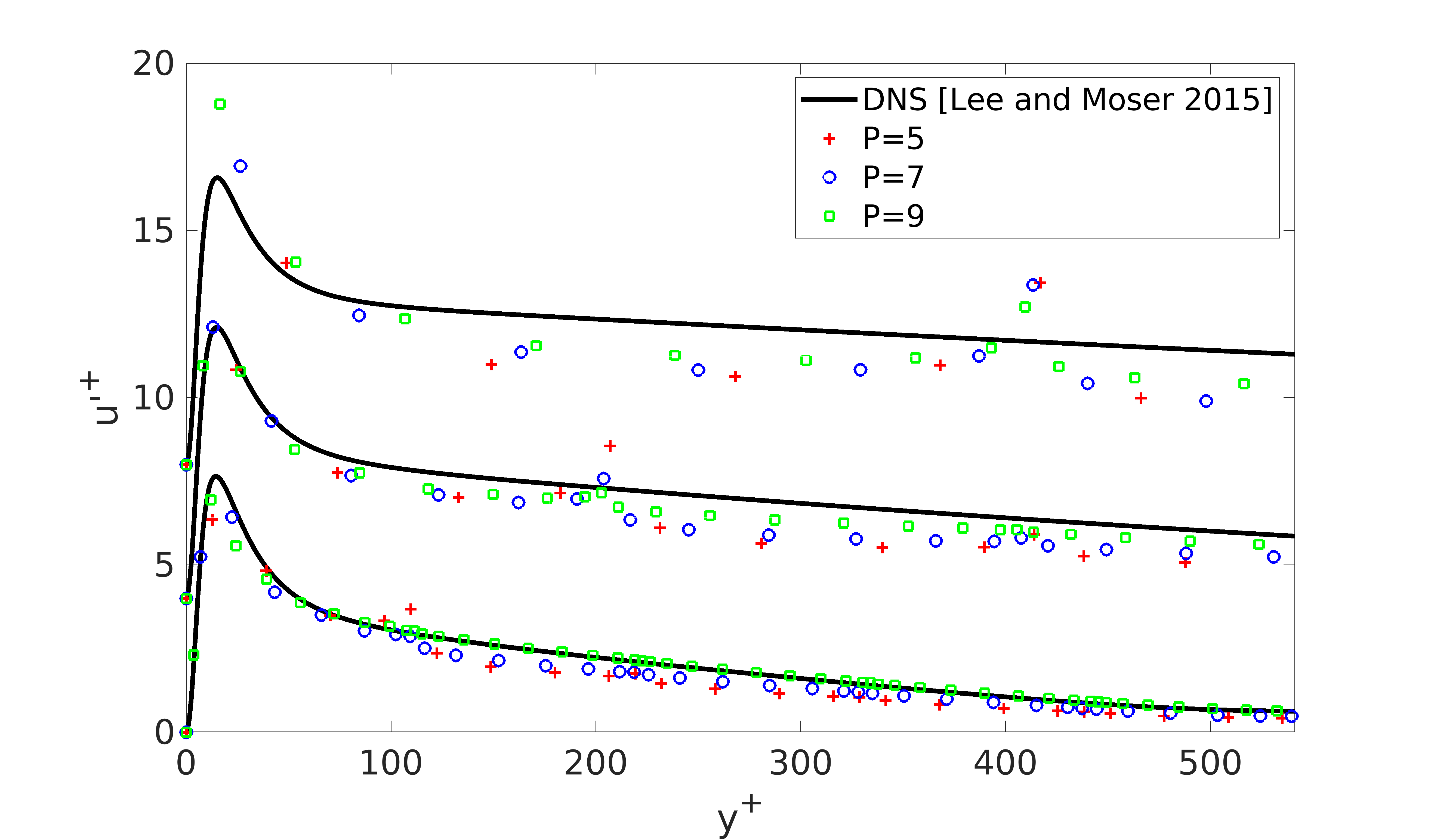}
     \end{subfigure}
     \hfill
     \begin{subfigure}[b]{\rmssize\textwidth}
         \centering
         \includegraphics[width=\textwidth]{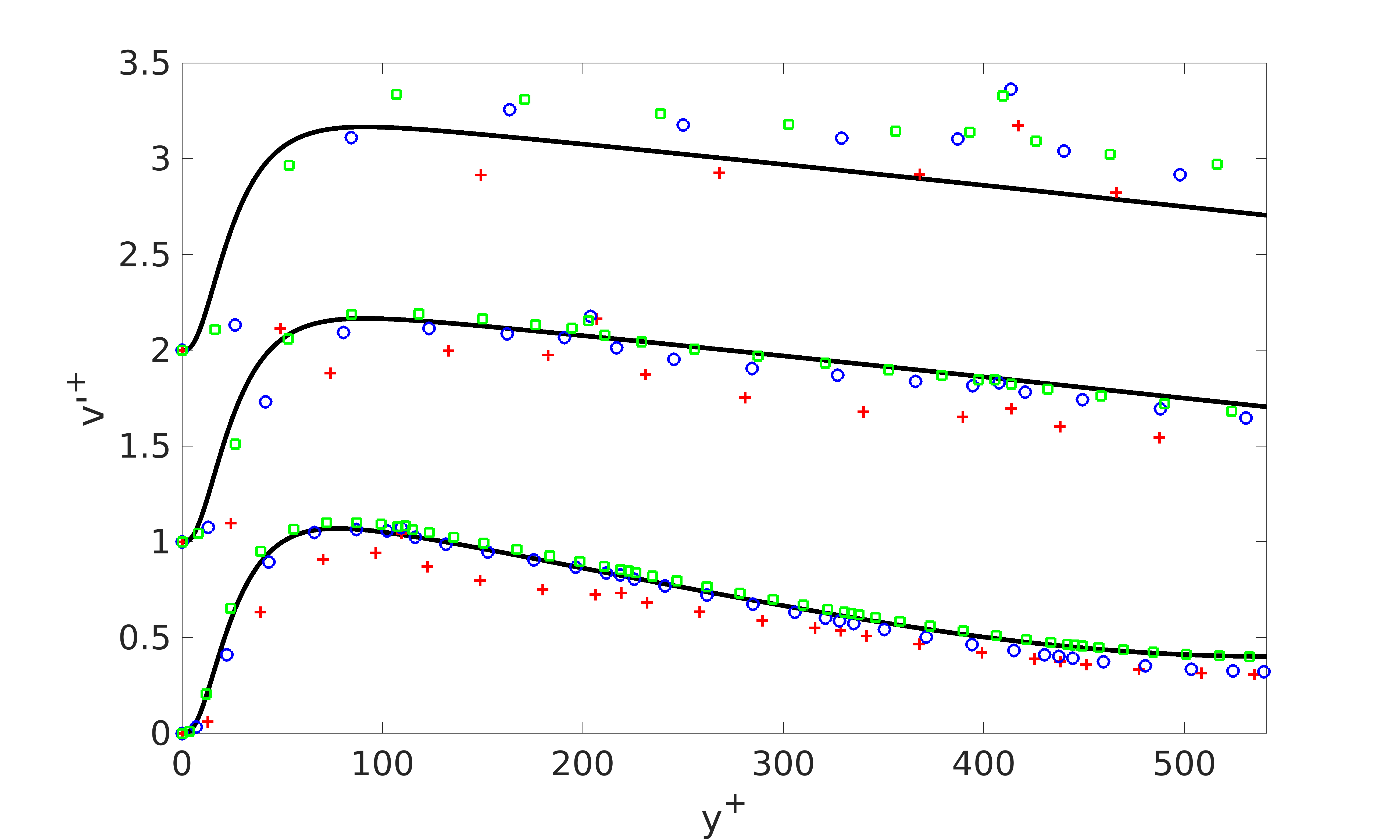}
     \end{subfigure}
     \hfill
     \begin{subfigure}[b]{\rmssize\textwidth}
         \centering
         \includegraphics[width=\textwidth]{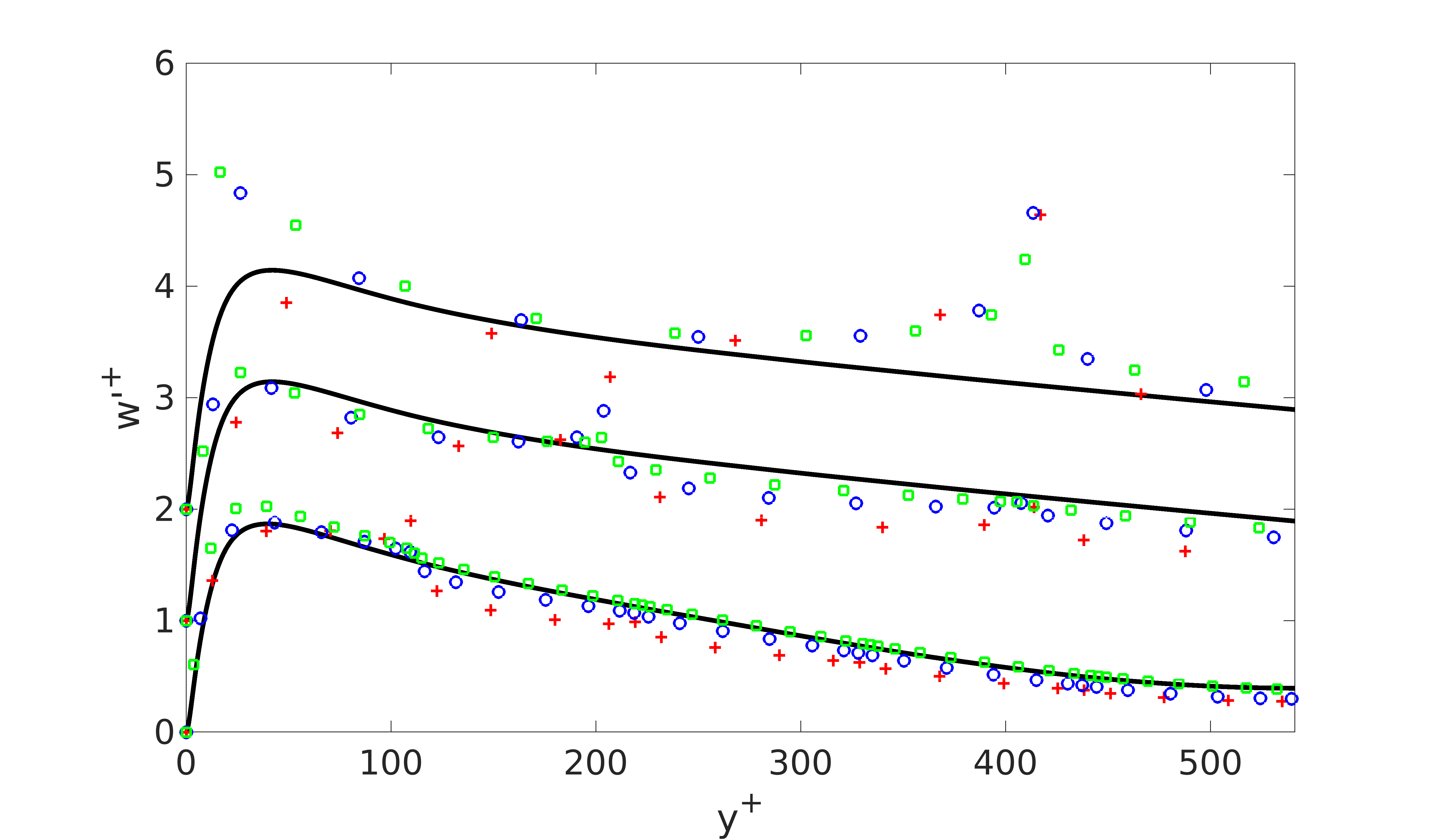}
     \end{subfigure}
     \hfill
     \begin{subfigure}[b]{\rmssize\textwidth}
         \centering
         \includegraphics[width=\textwidth]{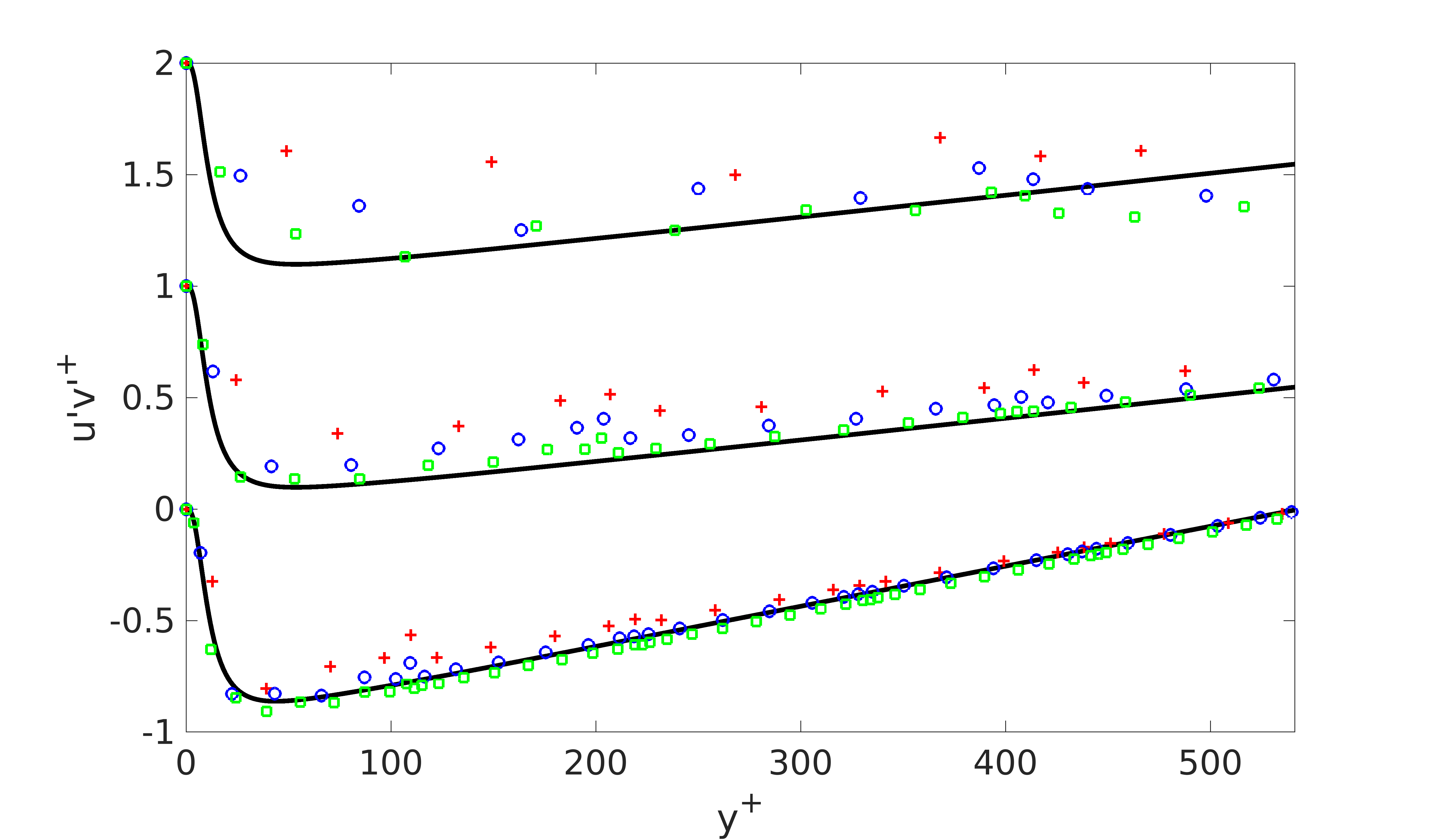}
     \end{subfigure}
	 \caption{Rms from WMLES of turbulent channel flow with $P=5,7,9$ order basis polynomials. Results for $Re_\tau=543$, $Re_\tau=1000$, and $Re_\tau=2000$, with $N=10$ are shown from bottom to top, shifted up by 4 for $u'^+$ and 1 for the other quantities for visibility. The solutions use a law-of-the-wall enrichment function in the wall-adjacent elements as the wall-model. The WMLES results are compared with the DNS data from~\cite{moser2015}.}
     \label{fig:rms_conv10}
\end{figure}

\begin{figure}
     \centering
     \begin{subfigure}[b]{\rmssize\textwidth}
         \centering
         \includegraphics[width=\textwidth]{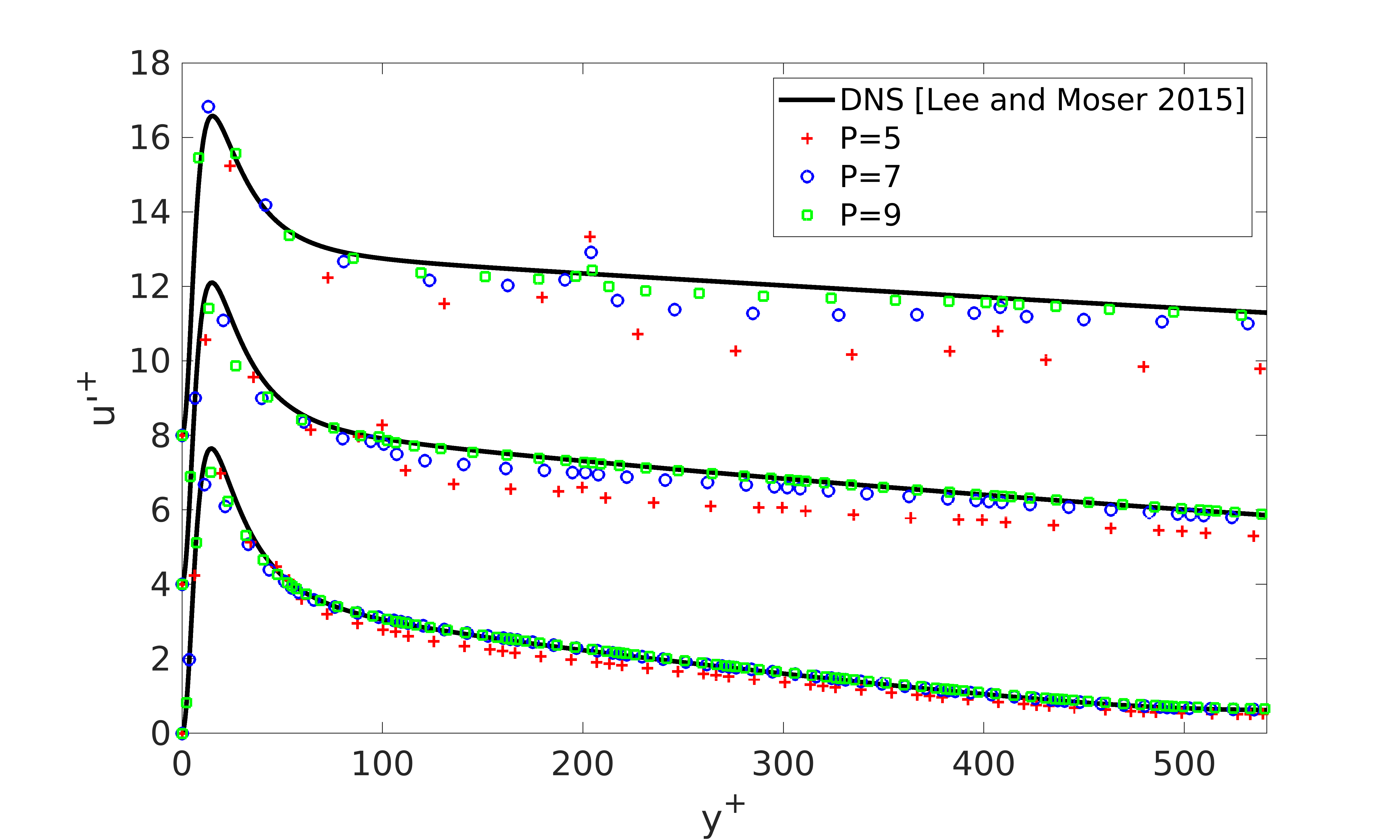}
     \end{subfigure}
     \hfill
     \begin{subfigure}[b]{\rmssize\textwidth}
         \centering
         \includegraphics[width=\textwidth]{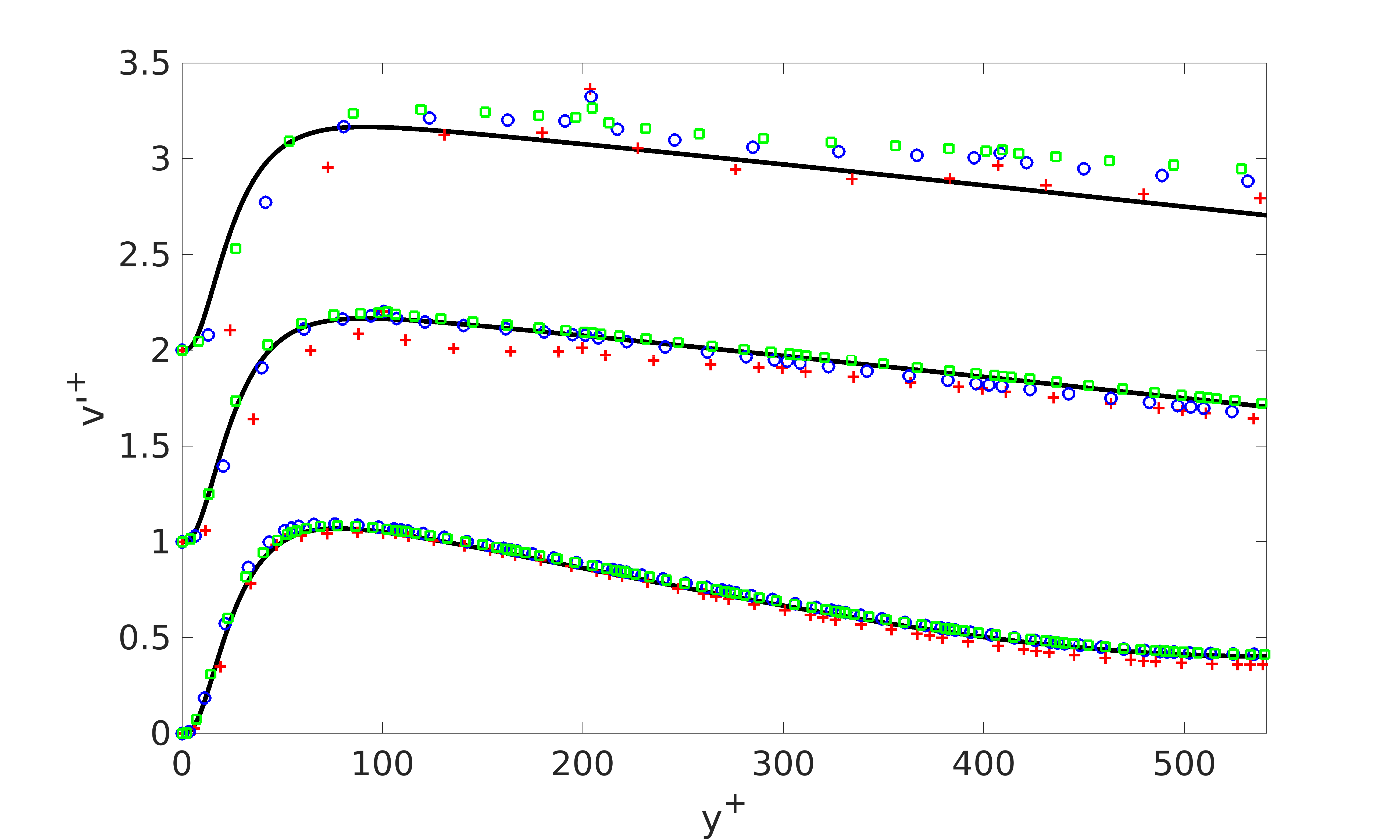}
     \end{subfigure}
     \hfill
     \begin{subfigure}[b]{\rmssize\textwidth}
         \centering
         \includegraphics[width=\textwidth]{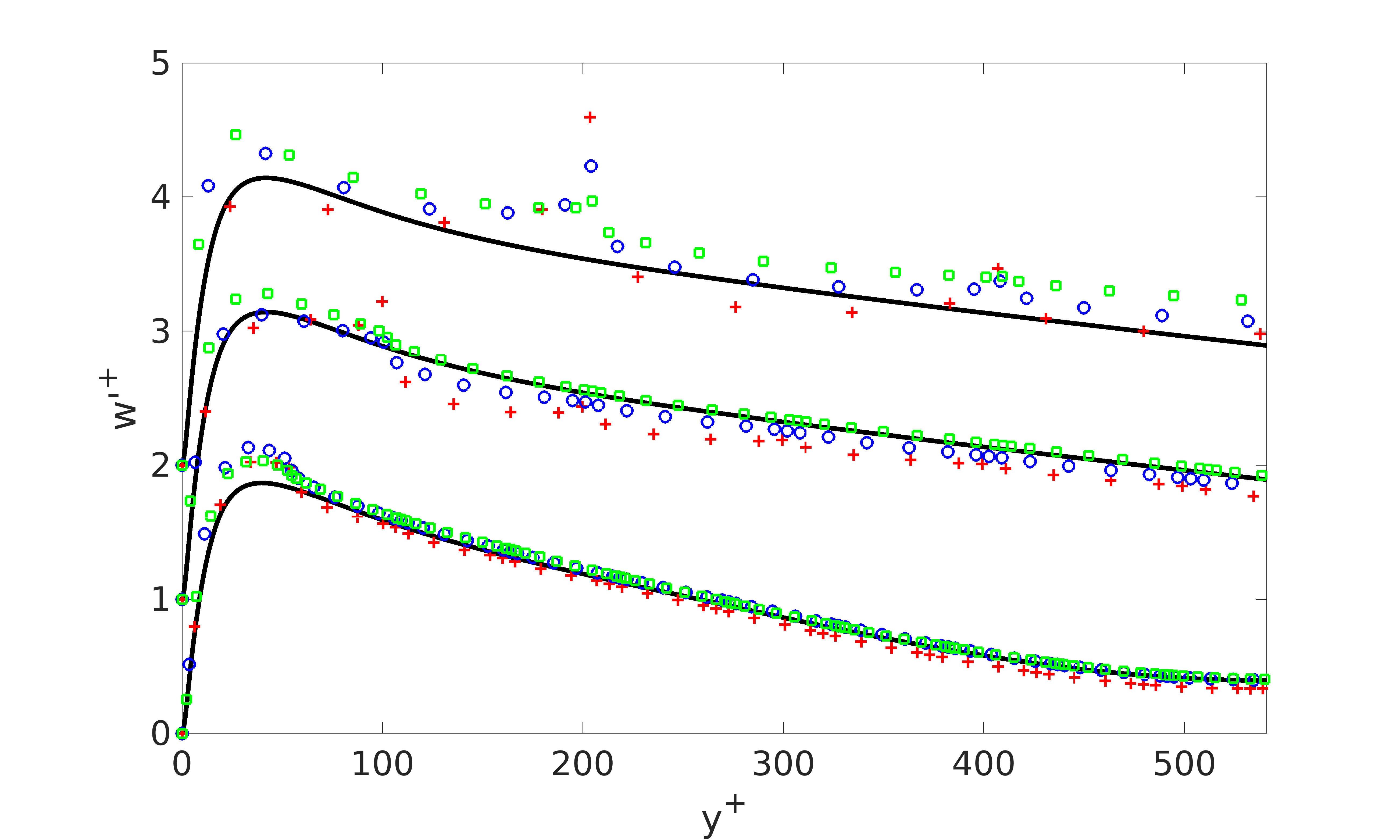}
     \end{subfigure}
     \hfill
     \begin{subfigure}[b]{\rmssize\textwidth}
         \centering
         \includegraphics[width=\textwidth]{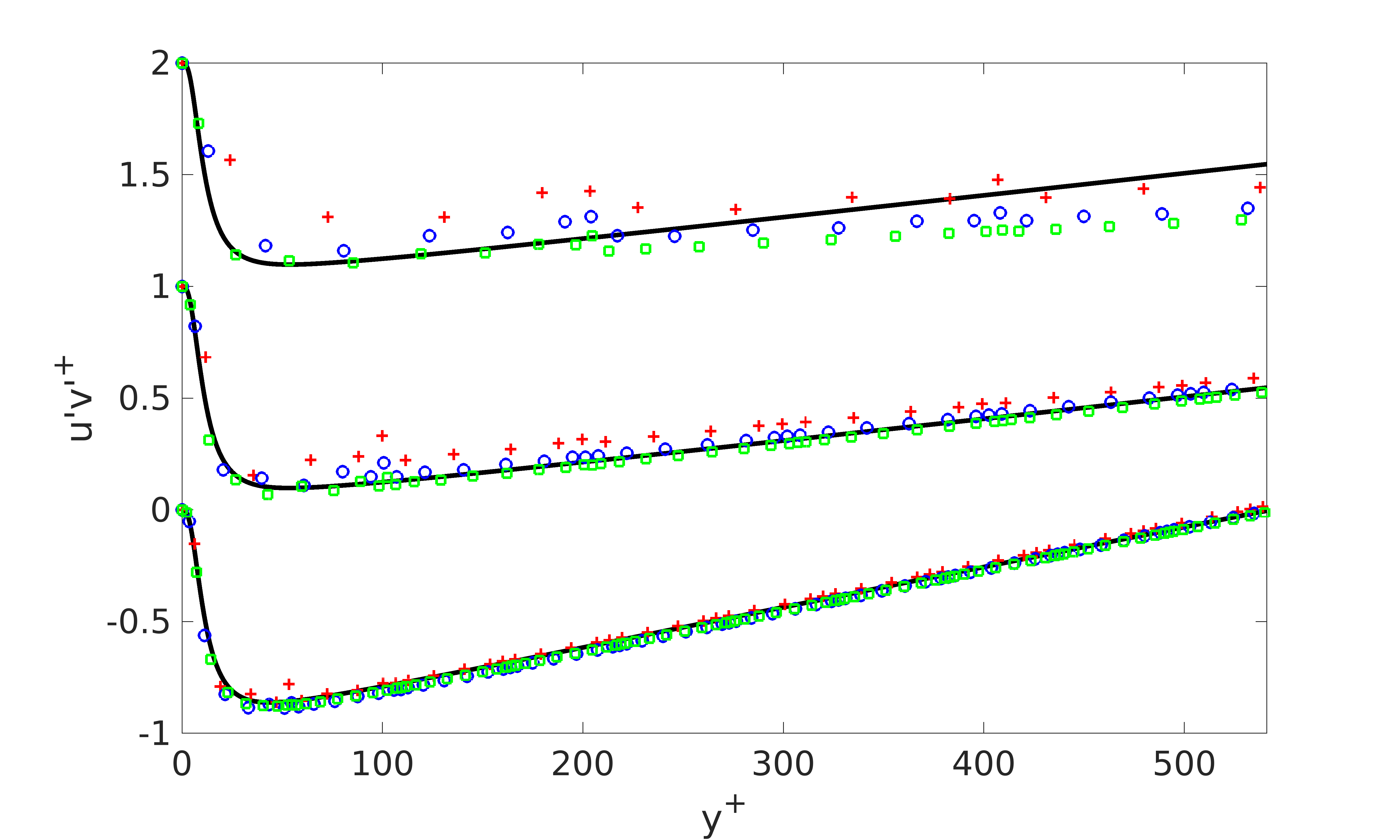}
     \end{subfigure}
	 \caption{Rms from WMLES of turbulent channel flow with $P=5,7,9$ order basis polynomials. Results for $Re_\tau=543$, $Re_\tau=1000$, and $Re_\tau=2000$, with $N=20$ are shown from bottom to top, shifted up by 4 for $u'^+$ and 1 for the other quantities for visibility. The solutions use a law-of-the-wall enrichment function in the wall-adjacent elements as the wall-model. The WMLES results are compared with the DNS data from~\cite{moser2015}.}
     \label{fig:rms_conv20}
\end{figure}

Figures~\ref{fig:mean_conv10}-\ref{fig:rms_conv20} show the mean velocity and rms profiles for turbulent channel flows with $Re_\tau=543$, $1000$, and $2000$ with $N=10$ and $N=20$ elements respectively. Table~\ref{tab:conv_y1} lists the size of the wall-adjacent element for the channel flow tests with $Re_\tau=543$, $1000$, and $2000$. The wall-adjacent elements are hundreds of plus units long and extend well into the bulk of the flow. The use of such large elements is enabled by the the enrichment wall-model. The results are shown for bases of order $P=5$, $7$, and $9$. The mean profiles are shifted up by $4$ for visibility as the Reynolds number is increased. For all of the cases, the mean velocity profile is in good agreement with the DNS data all the way up to the wall. There is no mismatch in the near-wall region because the no-slip boundary is enforced. The $Re_\tau=2000$ with $N=10$ shows some log layer mismatch, but as $N$ is increased to $20$, good agreement is seen. The rms data show similar trends to the DNS data. For the more underresolved cases, the rms values are underpredicted, however, as $P$ and $N$ are increased, better agreement is seen. In particular, the $N=20$, $P=9$ cases for $Re_\tau=543$ and $1000$ show good agreement in the rms data. As expected, better agreement with the DNS data is seen for the lower Reynolds number cases and cases with higher resolution. This indicates that the wall-model is helping the mean velocity profile match the desired flow conditions, while the polynomial SEM solution is capturing the second-order turbulent statistics outside of the boundary layer.

In the more underresolved cases, cusps in the rms data can be seen at the end of the first element. For example, for $Re_\tau=2000$, $N=20$, rms data cusps can be seen at $y^+=200$. This location corresponds with the end of the wall-adjacent element, which is the matching point location. The cusps are a result of the discontinuous gradients between the elements because only $C^0$ continuity is enforced at element boundaries. This behavior is not specific to the enrichment method, but is commonly seen in under-resolved high-order simulations \cite{fang2011towards,gillyns2022implementation,mukha2024wall}. It is a sign of an under-resolved mesh \cite{fang2011towards}, which is expected because the large wall-adjacent element motivates the need for the wall model. Figures~\ref{fig:rms_conv10} and \ref{fig:rms_conv20} show that the cusps can be remedied with increased resolution, like increasing $P$. Despite this inaccuracy in the gradients, good agreement in the mean velocity profile is seen, highlighting the strength of the enrichment wall-model. 

For all three $Re_\tau$ values, increasing $P$ and $N$ improves the agreement with the DNS as expected. Table~\ref{tab:conv_y1} shows that large elements are being used near the wall in all cases. For all three Reynolds numbers, resolutions with the first element around $y_1^+=200$ show good agreement with the mean but under-predicted rms quantities. However, for $y_1^+\leq 100$ both the mean and rms quantities show good agreement with the DNS data. If $N$ was refined such that $y_1^+<10$, the wall-model would have diminishing returns because the high-order polynomial solution will be more effective at capturing the near-wall region that does not include the log-layer. Larger values of $y_1^+$ lead to log-layer mismatch, as seen in the cases with $Re_\tau=2000$ and $N=10$. Hence, refining $N$ leads to more improvement than refining $P$, especially for the turbulent fluctuations. However, for a fixed mesh, increasing $P$ from $7$ to $9$ shows improvement in the RMS quantities, because the high-order polynomial can better represent the fluctuations near the wall. The larger Reynolds number cases require more refinement than the lower because of the larger gradients in the boundary layer profile. Overall, the enrichment wall-model effectively represents the near-wall behavior of the turbulent flow with large elements for a range of Reynolds numbers.

\begin{table}[]
\centering
\begin{tabular}{|c | c | c | c|}
\hline
 & $Re_\tau=543$ & $Re_\tau=1000$ & $Re_\tau=2000$  \\
 \hline
Case & $y_1^+$ & $y_1^+$ & $y_1^+$ \\
\hline
N=10 P=5 & 109.55 & 206.89 & 417.06 \\
N=10 P=7 & 109.33 & 203.77 & 413.46 \\
N=10 P=9 & 111.32 & 202.72 & 409.46 \\
N=20 P=5 & 54.28 & 99.80  & 203.51 \\
N=20 P=7 & 54.60 & 100.63 & 204.09  \\
N=20 P=9 & 54.72 & 102.37 & 204.60  \\
\hline
\end{tabular}
\caption{Wall-adjacent element size, $y_1^+$, for WMLES with the enrichment wall-model. }
\label{tab:conv_y1}
\end{table}

\subsection{Separation of Mean and Fluctuations}
\label{sec:mean_fluc}
\begin{figure}
     \centering
     \begin{subfigure}[b]{\rmssize\textwidth}
         \centering
         \includegraphics[width=\textwidth]{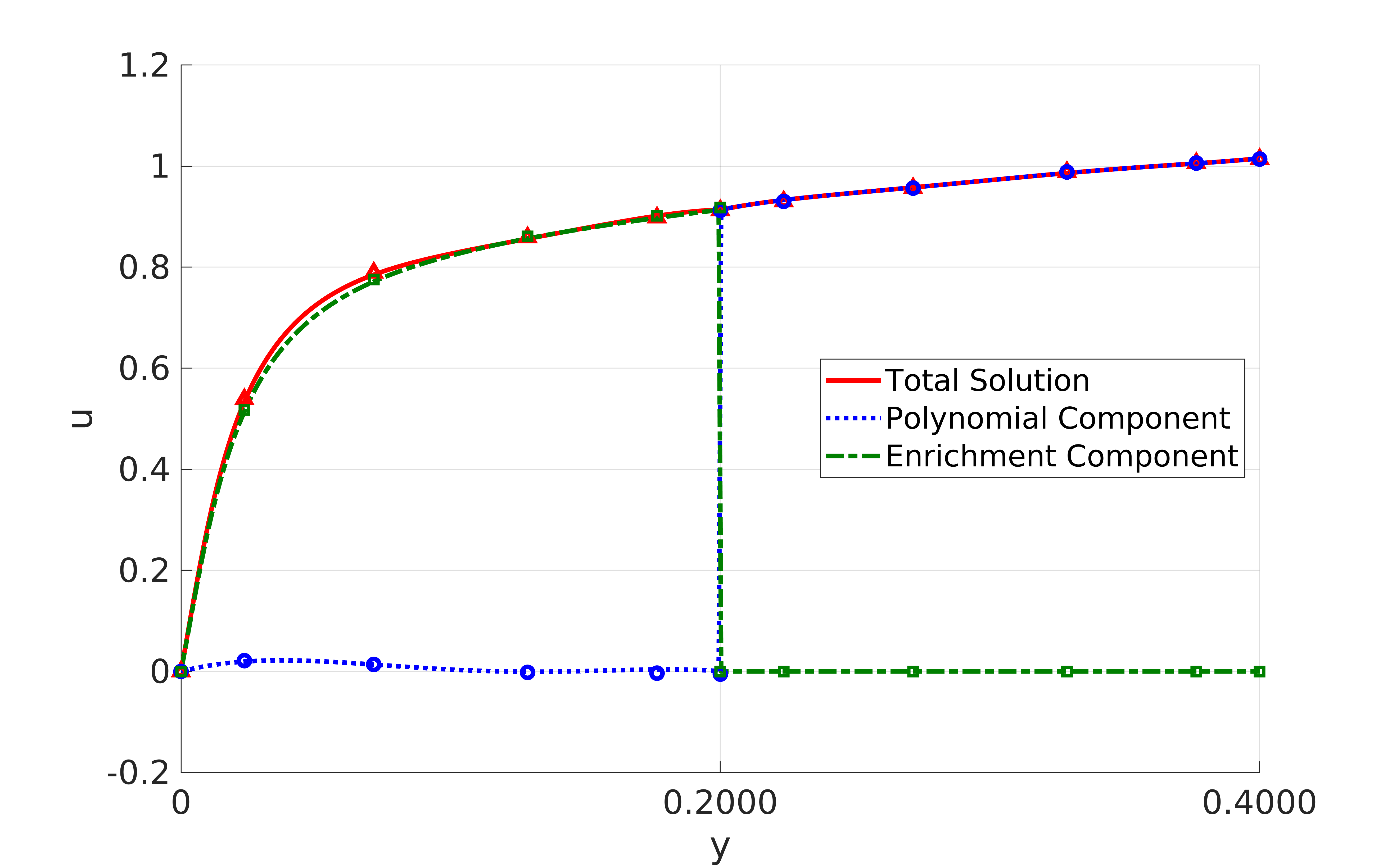}
     \end{subfigure}
     \hfill
     \begin{subfigure}[b]{\rmssize\textwidth}
         \centering
         \includegraphics[width=\textwidth]{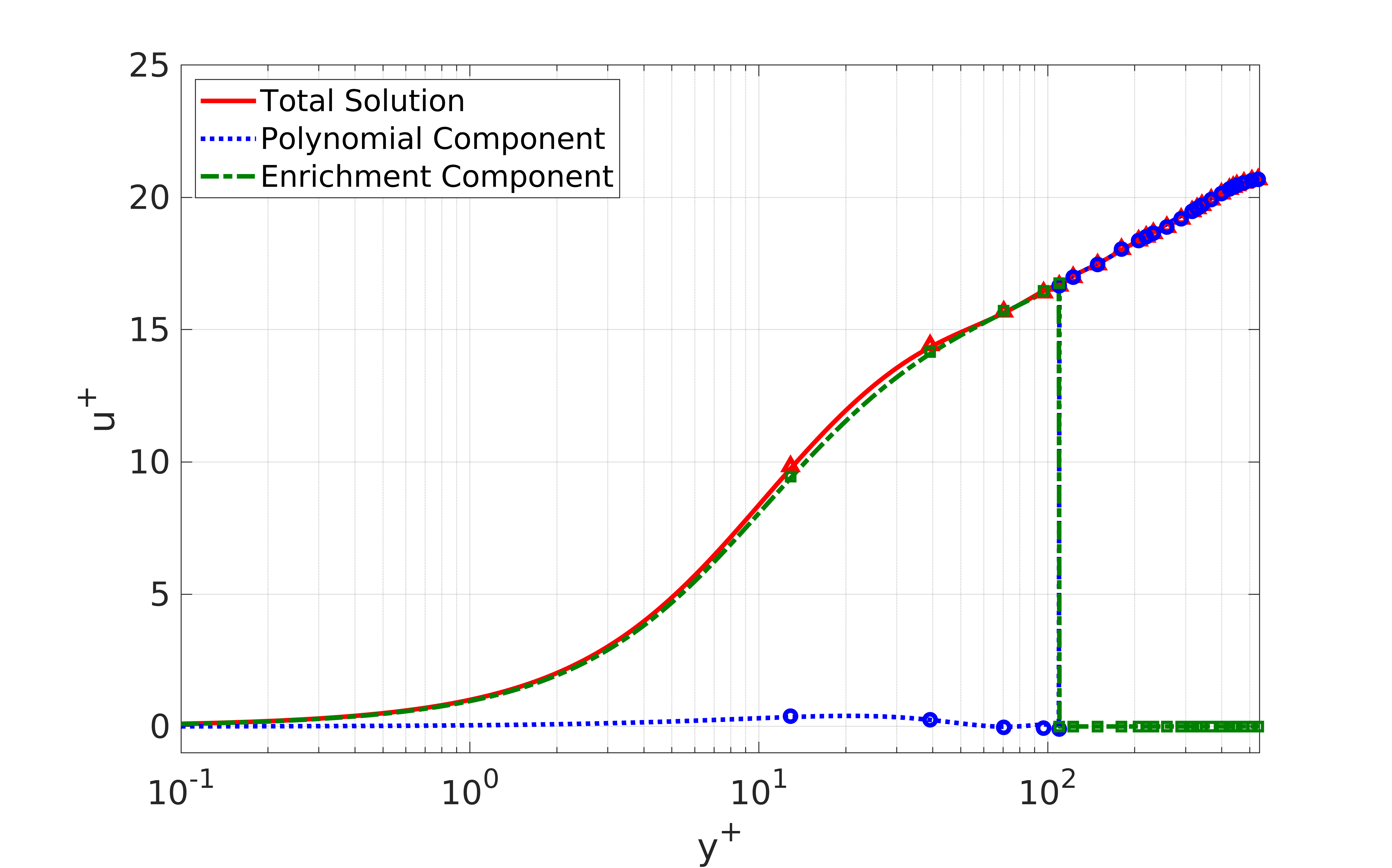}
     \end{subfigure}
     \hfill
     \begin{subfigure}[b]{\rmssize\textwidth}
         \centering
         \includegraphics[width=\textwidth]{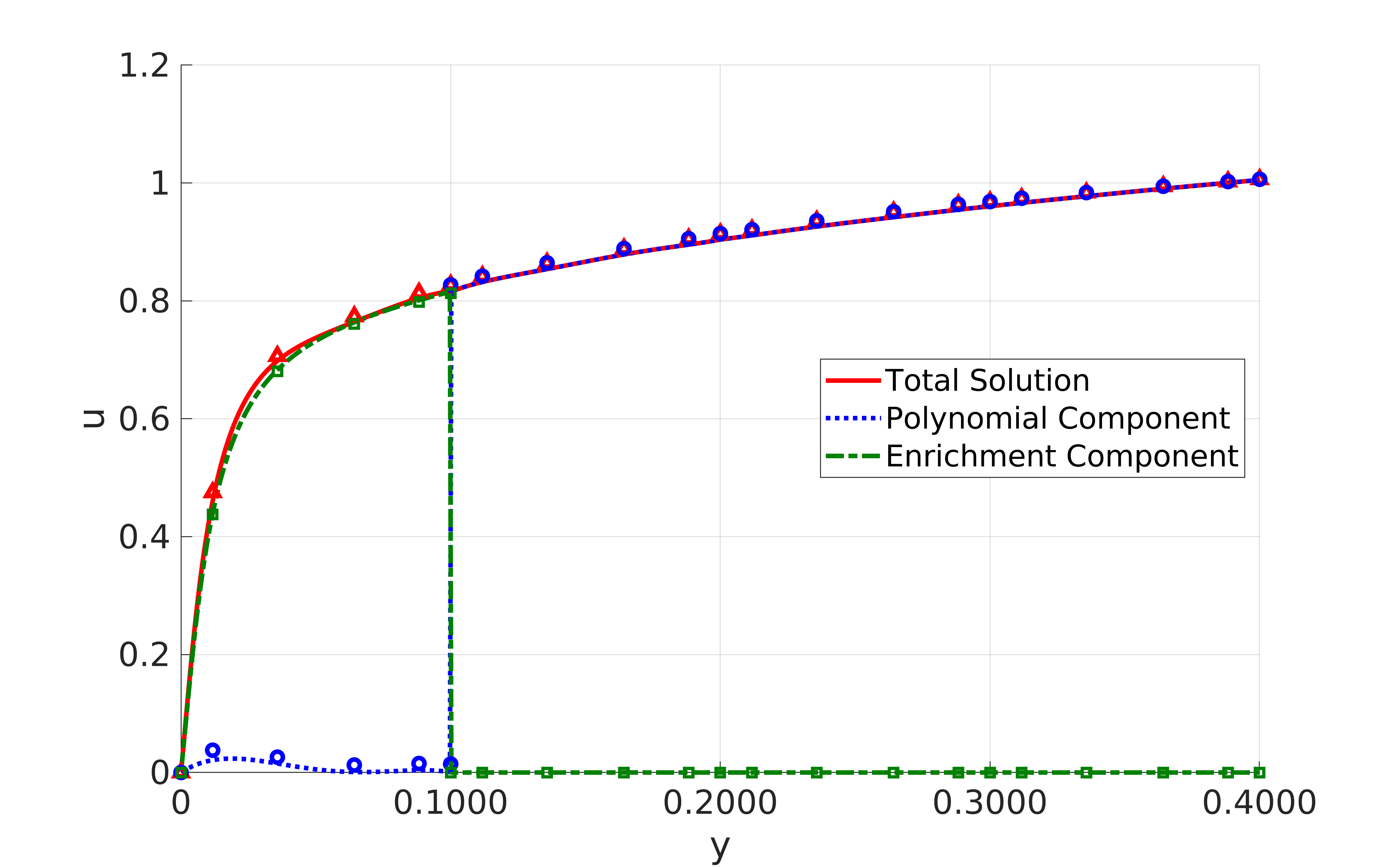}
     \end{subfigure}
     \hfill
     \begin{subfigure}[b]{\rmssize\textwidth}
         \centering
         \includegraphics[width=\textwidth]{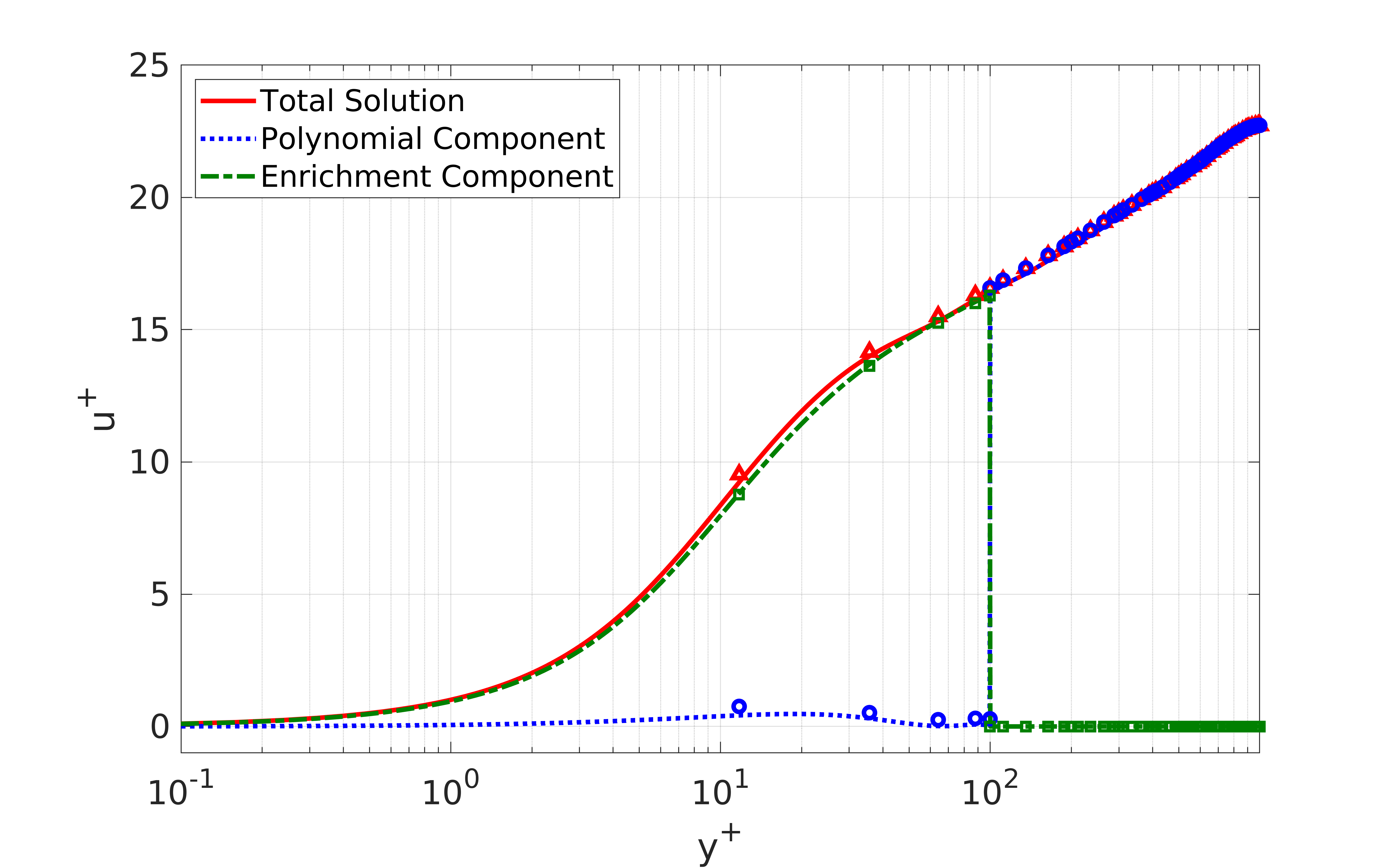}
     \end{subfigure}
	 \caption{Plots of mean velocity in physical space and plus units broken up into the enrichment and polynomial components. The top plots are for the $Re_\tau=543$, $P=5$, $N=10$ case and the bottom plots are for the $Re_\tau=1000$, $P=5$, $N=20$ case. In both cases the mean is captured by the enrichment function and the polynomial component has a nearly zero mean from representing the turbulent fluctuations.}
     \label{fig:mean_flucts}
\end{figure}

One of the attractive properties of the enrichment wall-model is the separation of the mean and fluctuations in the solution representation. The enrichment function is designed to capture the mean behavior, while the polynomial component captures the turbulent fluctuations. Figure~\ref{fig:mean_flucts} shows the mean velocity from the $Re_\tau=543$, $P=5$, $N=10$ and $Re_\tau=1000$, $P=5$, $N=20$ cases broken up into the enrichment and polynomial components. The vertical line in the plot corresponds to the edge of the first element because it is where the enrichment component becomes zero. The enrichment function contains the large gradients in the wall-adjacent element as expected. The mean of the polynomial component in the enriched portion of the domain is close to zero, as is expected because it captures the fluctuations around the mean. The mean polynomial solution is not exactly zero, because the enrichment function contains the modeled mean velocity profile near the wall and not the true mean profile. Overall, the enrichment method delivers the desired separation between mean and turbulent fluctuations.

\subsection{Comparison with Shear Stress Wall-Model}
\label{sec:results_comp}
In this section, the enrichment wall-model is compared with purely polynomial standard SEM and a shear stress wall-modeled SEM to show the benefits. All three methods use the same initial conditions and problem specifications from Sec.~\ref{sec:num}. The standard SEM uses no-slip walls and no added filtering, so the numerics match the enriched case exactly. At the coarse near-wall resolutions presented, using a filter for added dissipation in the SEM leads to large oscillations and a mismatch of the targeted turbulent profile. The shear stress wall-model is implemented as described earlier in~\cite{pal2021development,pal2022near}. The shear stress wall-model uses no-penetration slip-wall boundary conditions that apply a modeled shear stress at the wall. The matching point is taken as the top of the first wall-adjacent element. The model uses the Reichardt law-of-the-wall as the equilibrium wall-stress model. The initial solution is allowed to develop for $8$ flow-through times before the shear stress wall-model is activated. Additionally, an explicit filter with weight $0.05$ and cutoff ratio $0.65$ are used to add dissipation and greatly decrease the amount of slip at the wall. This added dissipation serves as the subgrid scale model for this LES, as is often done in SEM \cite{gillyns2022implementation,fischer2001filter}.

\begin{figure}[]
\centering
\includegraphics[width=\meansize\textwidth,trim={0 0 0 0cm}]{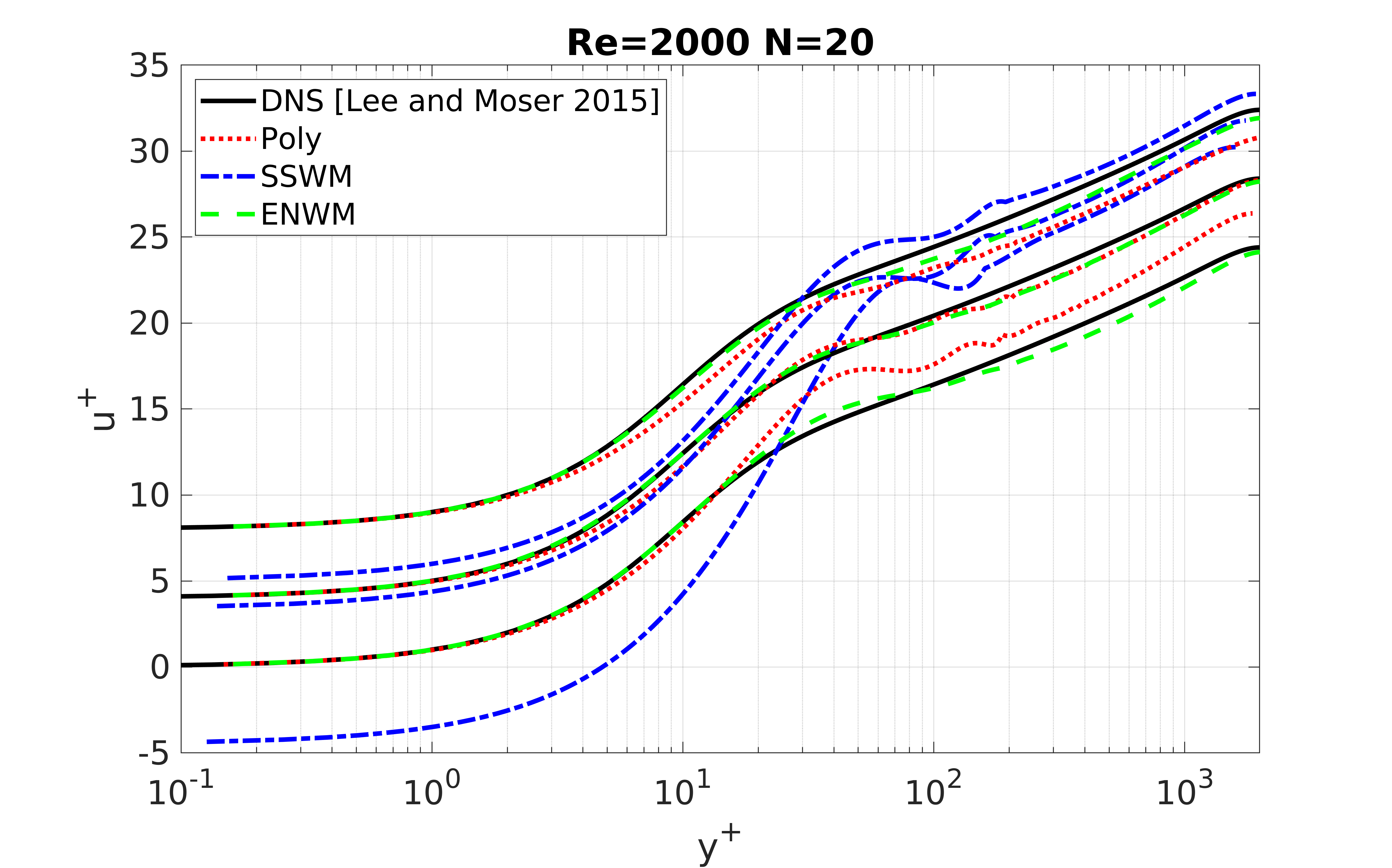}%
\caption{Mean velocity from $Re_{\tau}=2000$ with $N=20$ uniform elements in the channel height and $P=5,6,7$ order basis polynomials with standard SEM (Poly), shear stress wall-model (SSWM), and enrichment wall-model (ENWM). Results with different polynomial orders are shifted up for visibility with $P=5$ on the bottom and $P$ increasing as the results are shifted up by 4. The results are compared with the DNS data from~\cite{moser2015}.}
\label{fig:re2000_n20_poly_ss_en_ps}
\end{figure}

\begin{figure}
     \centering
     \begin{subfigure}[b]{\rmssize\textwidth}
         \centering
         \includegraphics[width=\textwidth]{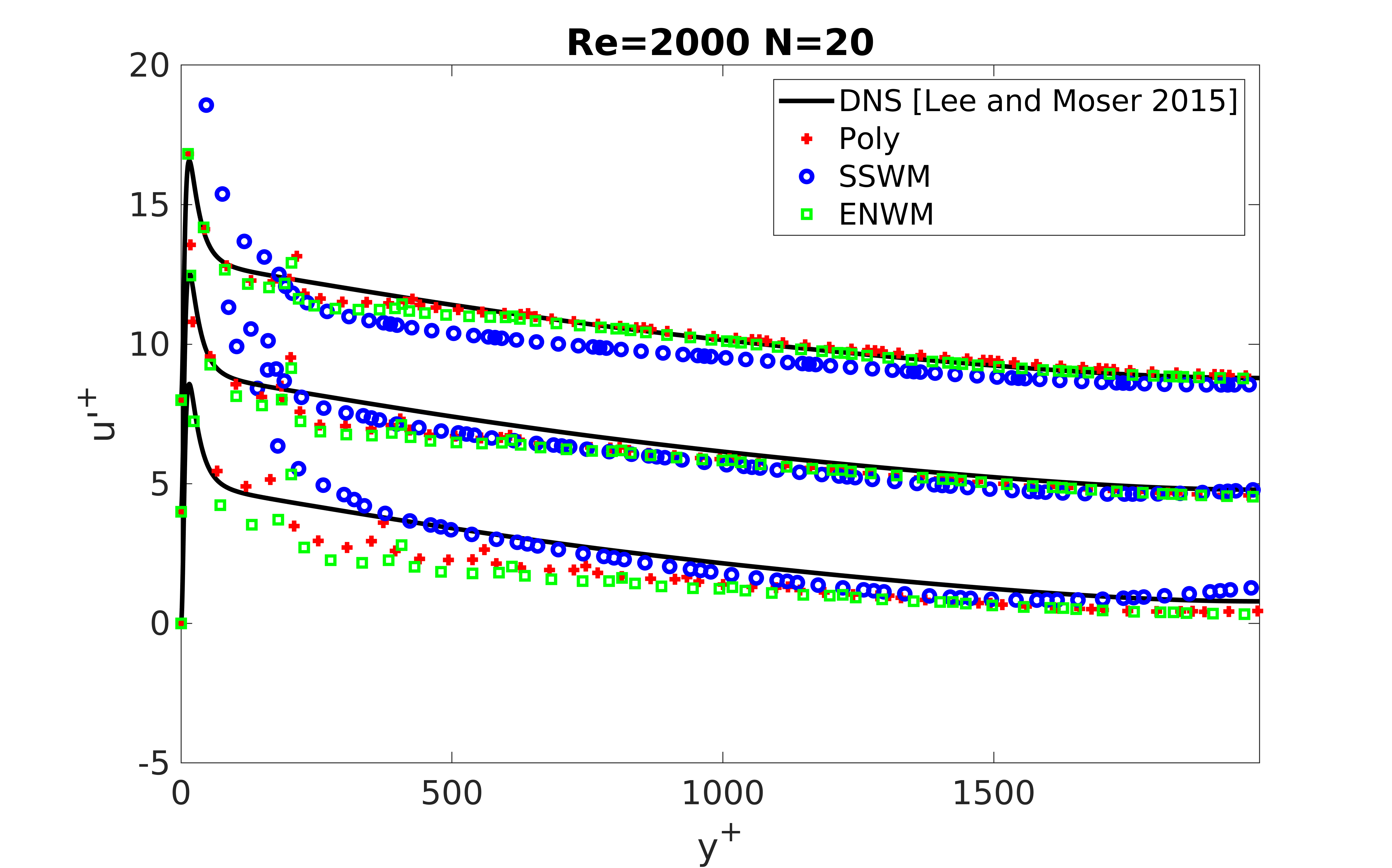}
     \end{subfigure}
     \hfill
     \begin{subfigure}[b]{\rmssize\textwidth}
         \centering
         \includegraphics[width=\textwidth]{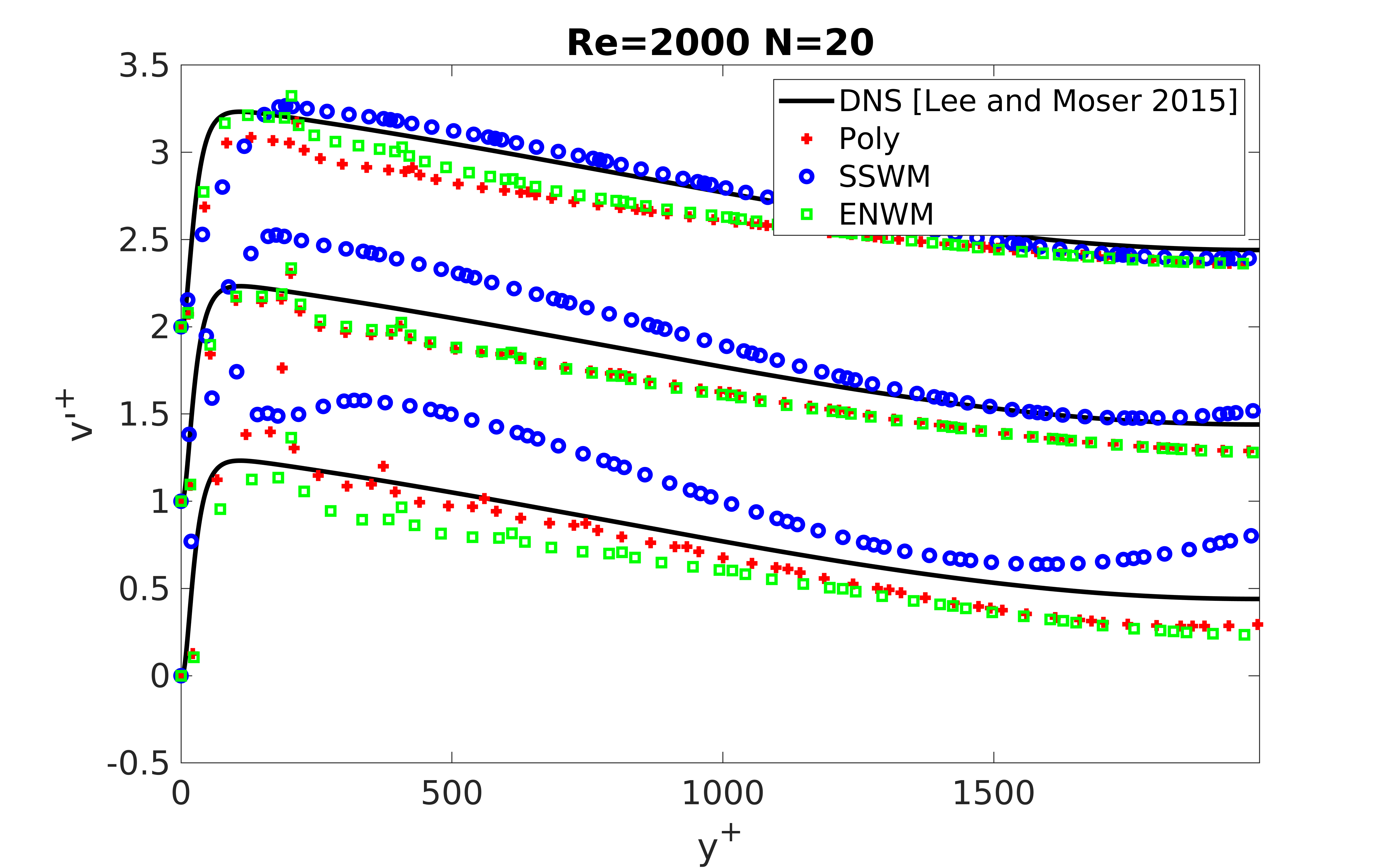}
     \end{subfigure}
     \hfill
     \begin{subfigure}[b]{\rmssize\textwidth}
         \centering
         \includegraphics[width=\textwidth]{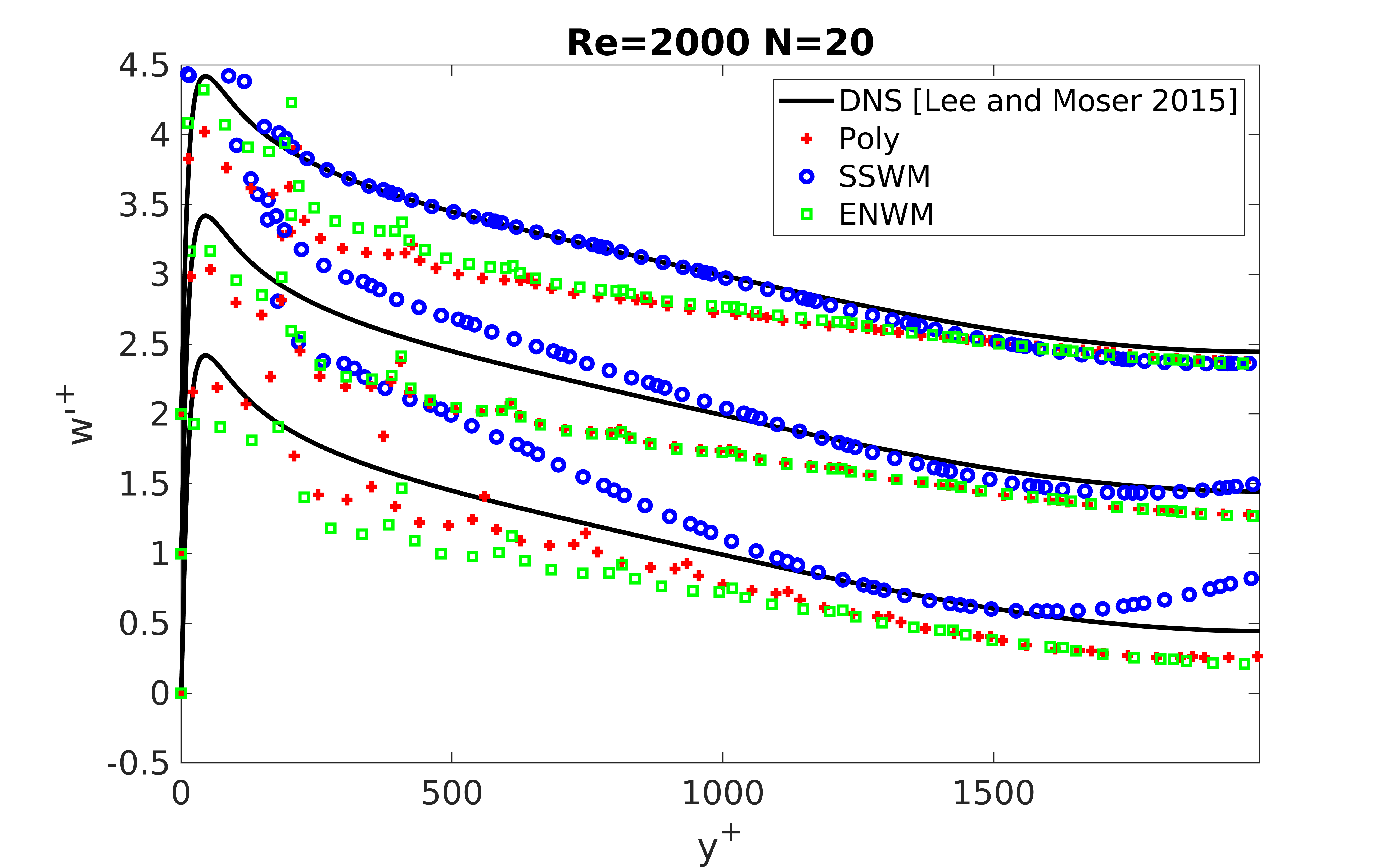}
     \end{subfigure}
     \hfill
     \begin{subfigure}[b]{\rmssize\textwidth}
         \centering
         \includegraphics[width=\textwidth]{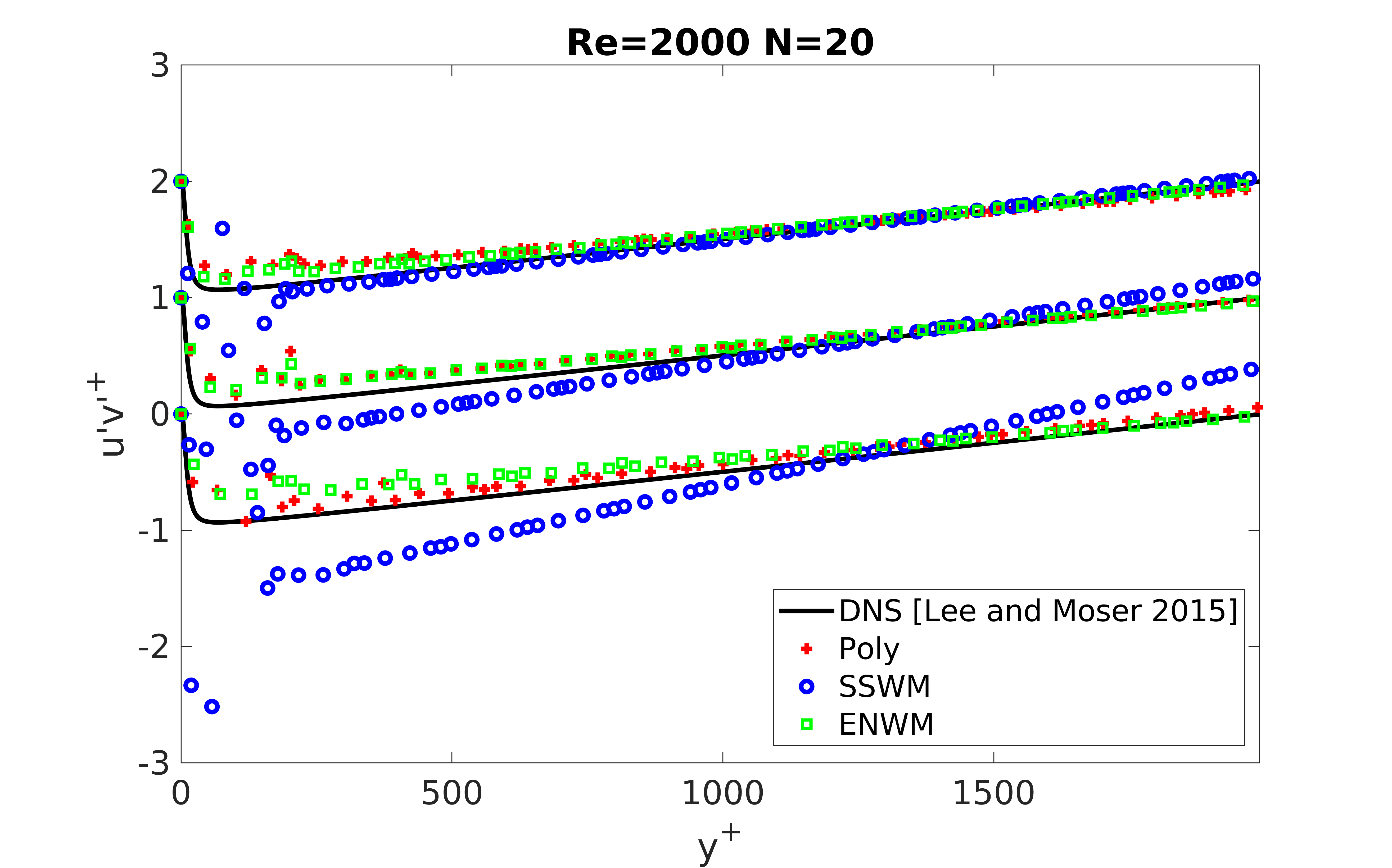}
     \end{subfigure}
	\caption{Rms quantities from $Re_{\tau}=2000$ with $N=20$ uniform elements in the channel height and $P=5,6,7$ order basis polynomials with standard SEM (Poly), shear stress wall-model (SSWM), and enrichment wall-model (ENWM). Results with different polynomial orders are shifted up for visibility with $P=5$ on the bottom and $P$ increasing as the results are shifted up by 4 for $u'^+$ and 1 for the other quantities. The results are compared with the DNS data from~\cite{moser2015}.}
     \label{fig:re2000_n20_poly_ss_en_ps_rms}
\end{figure}

Comparisons between the three methods for $Re_\tau=2000$, $N=20$, and $P=5,6,7$ are shown in Figures~\ref{fig:re2000_n20_poly_ss_en_ps}-\ref{fig:re2000_n20_poly_ss_en_ps_rms}. The standard SEM struggles to match the DNS mean velocity profile. The $P=5$ case displays oscillations in the first element and over-prediction in the log-layer, while the $P=6$ and $P=7$ cases under-predict the mean velocity. The shear stress wall-model has significant slip at the wall and oscillations in the first element. Then the mean profile is almost entirely missed in the log-layer. On the other hand, the mean velocity profile of the enrichment wall-model is much closer to the DNS data. The rms values show poor performance by the shear stress wall-model as it over-predicts most of the quantities and has oscillatory solutions. For these cases, the enrichment wall-model's rms data closely resembles the standard SEM. While the enrichment wall-model (ENWM) data improves with refinement in $P$, the method still under-predicts the rms values. In addition, the rms values in the wall-adjacent element are significantly different from the DNS data. For these $Re_\tau=2000$ cases, the enrichment wall-model greatly outperforms the other methods. It is able to achieve good agreement with the mean even for low resolutions where the standard SEM and shear stress model struggle. The enrichment model's no-slip wall is crucial to the accuracy of the mean velocity profile. 

Table~\ref{tab:utau_comp} shows the computed friction velocities for these cases. The enrichment wall-model predicts friction velocities within 2.5\% error, while the shear stress wall-model shows errors up to 20\% error. As seen in Figs.~\ref{fig:re2000_n20_poly_ss_en_ps}-\ref{fig:re2000_n20_poly_ss_en_ps_rms}, the standard SEM with $N=20$ and $P=6$ shows good agreement with the enrichment method and the DNS. However as $P$ increases, the friction velocity of standard SEM varies largely from the DNS data, showing that this method is not as converged as the enrichment method. The shear stress wall-model only applies the wall stress as a forcing term on the boundary condition and can only use the slip at the wall to prevent oscillations. Hence, for more under-resolved or higher $Re_\tau$ cases, the forcing term is not enough to ensure that $u_\tau$ is accurately predicted. The strength of the enrichment wall-model for predicting the friction velocity comes from the fact that the modeled wall stress is contained in the physical solution and the enrichment enables it to capture that profile without oscillations.

Comparing the enriched wall-model to the shear stress model and standard SEM demonstrates its strengths as well as some areas where it can be improved. The enrichment method shows significant improvement over the standard SEM in capturing the large gradients in the turbulent boundary layer and improves both the mean velocity profile as well as the rms quantities. Compared to the shear stress wall-model, the enrichment method shows better accuracy in capturing of the mean velocity profile in all regions. Even for the more resolved cases, the shear stress model only captures the log-layer profile and not the near wall region because of the slip at the wall. In the most under-resolved cases, the enrichment wall-model yields better agreement with the rms data than the shear stress model. However, in relatively more resolved cases, the enrichment method under-predicts the rms quantities. Hence, the enrichment wall-model shows improved prediction of the mean velocity, especially in very under-resolved situations, but there is scope for further improvement in capturing the rms quantities of interest which will be pursued in future work. In future work, we will assess the performance of the enrichment wall-model when various subgrid-scale turbulence modeling approaches are also included in the Nek5000 LES framework \cite{mukha2024wall}. In addition, the extension of the ENWM framework to non-equilibrium boundary layer flows will also be addressed in our future work.

\begin{table}[]
\centering
\begin{tabular}{|l | l | l | l | l | l | l|}
\hline
$Re_\tau$ & $N$ & $P$ & DNS $u_\tau$ & Poly $u_\tau$ & SSWM $u_\tau$ & ENWM $u_\tau$\\
\hline
2000 & 20 & 5 & 0.0459 & 0.0429 & 0.0368 & 0.0468 \\
2000 & 20 & 6 & 0.0459 & 0.0466 & 0.0404 & 0.0467 \\
2000 & 20 & 7 & 0.0459 & 0.0491 & 0.0444 & 0.0469 \\
\hline
\end{tabular}
\caption{Friction velocity, $u_\tau$ for standard SEM (Poly), shear stress wall-model(SSWM), and enrichment wall-model (ENWM) for $Re_\tau=2000$. The results are compared with the DNS data from~\cite{moser2015}.}
\label{tab:utau_comp}
\end{table}

\section{Conclusions}
In this work, we proposed the first enriched basis wall-modeled LES approach for the spectral element method. The enrichment framework augments the spectral element polynomial basis with a non-polynomial enrichment function. This enrichment function is an analytic law-of-the-wall function for the mean velocity profile in a turbulent boundary layer. Adding this function into the solution representation allows the enrichment to capture the large gradients while the polynomials capture the turbulent fluctuations. Additionally, the augmented solution accurately models the wall shear stresses without modification to the no-slip wall boundary conditions. As a result, the method can simulate high Reynolds number flows with large elements near the wall without unphysical oscillations. This model leverages the unique properties of the spectral element method to develop the enrichment framework and address the issue of unphysical regions in traditional wall-models.

The enrichment wall-model was validated on a range of turbulent channel flow cases. It was shown that the method can accurately capture the mean velocity profile with WMLES of channel flows up to the wall, without an unphysical region that occurs in shear stress wall-models. Compared to SEM with a shear stress wall-model, the enriched model shows a significant improvement in the mean velocity profile for under-resolved cases. There is some underprediction of the rms quantities in the enrichment method, which will be the focus of future work. However, the method shows improved agreement to the DNS mean velocity profile and rms data with increased resolution. The model also demonstrates a separability between the mean flow captured by the enrichment function and the fluctuations captured by the polynomial basis.

Future work will develop the method further for more challenging flows such as higher Reynolds numbers and flows with separations. For higher Reynolds number flows where $y_1^+<200$ is not possible, we plan to develop a dissipative filter compatible with the enrichment method to address the log-layer mismatch and under-prediction of rms quantities that come from buildup of energy in the large wavenumbers. For simulations with flow separation, more complex wall-functions, such as differential equation based wall-functions, can be directly incorporated into the enrichment framework. Additionally, a sensor will be developed to disable the enrichment in regions of flow separation because the polynomial solution and no-slip boundary conditions should be sufficient to resolve the flow.

\section{Acknowledgments}
This work was performed under the auspices of the U.S. Department of Energy by Lawrence Livermore National Laboratory under contract DEAC52-07NA27344. This article has been assigned an LLNL document release number LLNL-JRNL-860809-DRAFT. The submitted manuscript has been created by UChicago Argonne, LLC, Operator of Argonne National Laboratory (Argonne). Argonne, a U.S. Department of Energy Office of Science laboratory, is operated under Contract No. DEAC02-06CH11357. The U.S. Government retains for itself, and others acting on its behalf, a paid-up nonexclusive, irrevocable worldwide license in said article to reproduce, prepare derivative works, distribute copies to the public, and perform publicly and display publicly, by or on behalf of the Government. The research was funded by the Department of Defense (DoD) through the National Defense Science \& Engineering Graduate Fellowship (NDSEG) Program and the DOE project DE-EE0008875. Lastly, the authors would like to acknowledge the computing core hours available through the Bebop cluster provided by the Laboratory Computing Resource Center (LCRC) at Argonne National Laboratory for this research.

\bibliographystyle{unsrtnat}
\bibliography{newref}

\end{document}